\documentclass[aps,pra,reprint]{revtex4-1}
\usepackage{times}
\usepackage{graphicx}
\usepackage{bm}
\begin{document}


\title{Charge Conjugation, Heavy Ions, $e^+ e^-$ pairs: Was there a better way to add potentials to Dirac's free electrons?}




\author{Samuel P. Bowen}
 \affiliation{             Chicago State University}
               \email{sbowen@csu.edu}           
\author{            Jay D. Mancini}
 \affiliation{              Kingsborough Community College}

\date{Received: Mar. 23, 2012 }

\begin{abstract}
This is a study of a possible alternative procedure for adding a potential energy to the free electron
Dirac equation.  When Dirac added potentials to his free electron equation, there were two alternatives (here called D1 and D2).
He chose D1 and lost charge conjugation symmetry, found Ehrenfest equations that depended on the sign of the energy of the state determining the expectation value, encountered Klein tunneling, zitterbewegung and the Klein paradox.  The $D1$ alternative also predicted that deep potentials should pull positive energy states down into the negative energy continuum, possibly creating an unstable vacuum.  Extensive experiments (1975-1997) found no evidence for this instability, but did find low energy electron-positron pairs with sharply defined energies and unusually low counting statistics.  These pairs tended to disappear with higher beam currents.
This paper explores the other alternative, here called $D2$ and finds charge conjugation symmetry preserved, Ehrenfest equations are classical, Klein tunneling is not present, unstable vacuua are forbidden, zitterbewegung is absent in the charge current density, new excitations of bound electron-positron pairs are possible in atoms, and the energies at which low energy electron-positron pair production in heavy ion scattering occurs is well described. Also all of the positive energy calculations, including those with the Coulomb potential, the hydrogen-like atom, are retained exactly the same as found in alternative $D1$.  It might have been better if Dirac had chosen alternative $D2$.
\end{abstract}





\maketitle

\section{Introduction}

In the period 1973-1999 there was considerable activity seeking to use the attractive potential of two heavy ions scattering to pull down the lowest bound state electron states of the two nucleon system down through the mass gap $(-m,m)$ , (here $c=1$), into the negative continuum states.  The idea\cite{Greineretal1,Greineretal2,GandH} was to destabilize the Dirac Sea by bringing an electron state into contact with the negative energy states.
The possibility of making the vacuum unstable generated a great deal of theoretical and experimental excitement which was documented in three conference proceedings\cite{Landstein,Maratea86,Cargese90}.  The electronic Dirac equation seemed to predict fascinating physics if the positive energy bound states could be pulled down into the negative continuum.

However, after many experiments the only surprising observation\cite{Landstein,Maratea86,Cargese90} was the existence of a very narrowly defined total energy peak for an electron-positron pair at a low energy of hundreds of $KeVs$.  The best theory had predicted that there should be a dependence on the total charge $Z_1+Z_2$ of the two heavy nuclei and that there should be a critical threshold for this sum.  Above this critical sum there should be evidence of vacuum instability and nothing exceptional should be observed below this threshold.  In fact, the electron-positron sum energies for various heavy ion pairs did not correlate well with the nuclear charges and sub-critical pairs\cite{subcritical} also displayed the sharply peaked total energy $e^-e^+$ pairs.

The early experiments were capable of detecting only one member of the electron-positron pair, but as the experiments grew in sophistication, detection of both members of the pair was possible and the narrowness of the total energy distribution was verified. The mysterious source of these $e^-e^+$ was apparently generated in the collisions and moved at a slow speed from the scattering center.  The counts for these $e^-e^+$ pairs were generally not very large. They were difficult to detect and it took very long runs to generate good statistics.

A collaboration called APEX\cite{apexprop} was organized through Argonne National Laboratory to use
the ATLAS\cite{atlas} heavy ion accelerator, which was capable of large ion currents, and to use specially
developed spectrometers for detecting simultaneously the electrons and positrons from the collisions.  The hope was to use the larger currents to lower run times for generating adequate statistics.  The early runs\cite{Reviewtalk} at low beam currents generated fairly convincing data that a sharp total energy peak was present, though the counting statistics were low.

However, as the beam current was increased, the peak did not grow out of the background, rather it seemed to disappear as the run proceeded.  The properly skeptical experimentalists decided that the failure of the sharp peak to survive large beam currents and long running times indicated it was an artifact, an unexplained, unreliable chimera\cite{apexlet},\cite{apexlet2},\cite{apexarticle}.  Despite reasonable objections\cite{sciencelet}, the APEX results essentially closed down research in this whole area.  There have been almost no papers on this subject since 1999.

What could have gone wrong?  The Dirac equation for deep square wells predicts that bound state energy levels can be pulled below the bottom of the mass gap.  Yet no vacuum instability was observed.  None of a great variety of theoretical ideas could predict the low energy, sharply defined total energy $e^-e^+$ peaks.  Also none of the theories could explain either the wide spread observation of these $e^-e^+$ pairs for many heavy ion pairs at low beam currents or the disappearance of the sharp peak as the beam current increased.  What could be the reason that the efforts of so many capable scientists were not able to confirm the predictions and explain the sharp $e^-e^+$ peaks?

 A very large number of theoretical ideas have been explored hoping to explain these failures.  So far it appears that no one has gone back to the beginning of relativistic quantum mechanics and considered the process that Dirac used to add in the potentials to the free electron theory.  Before Dirac added in the potential energy, his free electron equation had charge conjugation invariance and the equations of motion of the expectation values obeyed the Ehrenfest classical equations.

The goal of this study is to re-examine Dirac's derivation\cite{Dirac1},\cite{Diracbook} of his
equations and the way he included a potential energy in his free electron equation. The
possibility that there may have been a alternative "path not
taken"\cite{pathnottaken} is explored and this possibility seems to recover much of Dirac's results,
resolves the above issues and also points to new physics that may have been
excluded by Dirac's derivation.\ \ In the following, the path taken
by Dirac will be labeled $D1$ and the other alternative will be labeled $D2.$
\ Before delineating these alternatives, it is useful to summarize some of the
theoretical consequences of alternative $D1.$

\subsection{Consequences of Dirac's Choice for Incorporating Potential Energy}

The non-interacting Dirac equation has an energy spectrum symmetric about zero
energy and the equation obeys charge conjugation invariance\cite{Merzbacher},\cite{BandD}.  That is, the negative
energy solutions are mapped onto the positive energy solutions and vice versa.
\ Yet, when a 4-vector potential $A_{\mu}$ is included using alternative $D1$, the charge conjugation
invariance breaks down and the energy spectrum is no longer symmetric about zero energy.

 The sign of the vector potential is seen to change as the charge conjugation transformation is applied to solutions of the interacting Dirac equation.  This has long been interpreted to mean that the positrons, which are represented by the transformed wave functions $\psi_c$ have the opposite charge to that of the electrons\cite{Diracbook},\cite{Merzbacher},\cite{BandD}.  However, the sign of the positron was originally assigned by looking at the change in the charge of the Dirac vacuum in the absence of a negative electron.  This assignment of charge is valid even in the non-interacting Dirac equation where there is no vector potential or explicit charge in the Hamiltonian.  This observation would support the argument that the sign change in the vector potential in the transformed Dirac equation could simply be interpreted as evidence that the charge conjugation invariance of the Dirac equation breaks down.  The interpretation of the charge of a hole in the vacuum and its positive energy relative to the vacuum is independent of the charge conjugation transformation itself.

In the $D1$ alternative, Dirac's original approach, which so strongly informs our intuition, we have an apparently consistent theory of electrons and positrons, even though there are some inconsistencies: (1) failing to be charge conjugation invariant in the presence of a vector potential, (2) having Ehrenfest equations of motion for expectation values that depend on the sign of the energy of the wave function forming the expectation value, and (3) having the existence of Klein tunneling in which low energy electrons appear to tunnel into high and wide potential steps that are higher than the mass gap of $2m$.  Finally, the perplexing failure of the heavy ion scattering program to discover an unstable vacuum and the appearance of narrowly defined total energy of $e^-e^+$ pairs, which seem to disappear at high beam currents, suggest that the $D1$ alternative could be inconsistent with nature.

The Ehrenfest theorem, that the equations of motion of expectation values of
various operators should reproduce the classical equations of motion\cite{Merzbacher,BandD}, has long been a limiting check on the validity of quantum
mechanical theories. \ Derivations in the alternative $D1$ of the equation of motion for the canonical
momentum for a particle (electrons with negative charge) in an electric and
magnetic field from the interacting Dirac equation ends up with the sign of
the Lorentz force law depending on the sign of the energy of the state being
used to determine the expectation value of the canonical
momentum\cite{Merzbacher}.

A similar $D1$ derivation from the interacting Dirac equation of the motion of a
spin in a magnetic field shows that the direction of the torque on the spin
again depends on the sign of the energy of the state being used to calculate
the expectation of the spin components\cite{Merzbacher}. \ Both of these results contradict the Ehrenfest theorem that
the equations of motion for the expectation values should be classical.

Similarly, the probability current density in $D1$ should be a constant in the limit of no electric field, but
it remains time dependent in the free particle limit while the momentum remains constant.
This also represents a violation of the classical limit of an Ehrenfest equation.  The probability
current density also exhibits zitterbewegung oscillations at high frequencies of $2m$.

The Klein paradox\cite{BandD},\cite{Klein} has disclosed a further
complication with the interacting Dirac equation in the $D1$ alternative. \ There have been a wide
variety of demonstrations that something strange occurs if a potential step is
too large in energy.  Recently, Dragoman\cite{Dragoman} and Bowen\cite{Bowen} have noted
that the traditional view of the Klein paradox, that the
reflection and transmission coefficients are not positive, is
easily resolved analytically, but the Klein tunneling remains unresolved.

Finally, combining these theoretical observations of the breakdown of charge conjugation invariance,
defective Ehrenfest equations of motion, and Klein tunneling, with the experimental failure to
observe a straight forward prediction of vacuum instability and the lack of a theory for the sharp
$e^-e^+$ pairs, argues that a second look at Dirac's procedure for adding in the potentials should
be undertaken.

The rest of the paper is organized as follows: \ In section II the
two alternative approaches to relativistic quantum mechanics will be delineated with the following:
\begin{enumerate}
\item Review of Dirac's free particle equation and its solutions.
\item Addition of the potential energy to the free electron equation and description of the $D1$ and $D2$ alternatives.
\end{enumerate}
Section III presents several topics comparing the alternatives $D1$ and $D2$ in the following sub-sections.
\begin{enumerate}
\item Charge conjugation invariance survives the addition of vector potentials n $D2$, while it fails in $D1$.
\item Ehrenfest equations deficiencies in $D1$ and resolution of these in $D2$.
\item Effect of different one electron states used in $D1$ and $D2$.
\item Bound states in the mass gap in $D1$ and $D2$.
\item Differences between $D1$ and $D2$ for piecewise constant potentials.
\item Klein tunneling in $D1$ and $D2$.
\item Proof by contradiction of a limit for attractive potentials in $D2$.
\item Aspects of Dirac hole theory in both alternatives.
\item Probabililty current density and zitterbewegung in $D1$.
\item Charge current density and the absence of zitterbewegung in $D2$.
\item Completeness in $D2$.
\item Transformations and Feynman Stuckleberg theory in $D1$ and $D2$.
\item Feynman diagrams and Perturbation Theory in $D2$.
\end{enumerate}

Section IV compares the consequences of $D2$ and Heavy Ion Experimental Data with the following subsections:
\begin{enumerate}
\item Historical Introduction
\item Hydrogenic Bound states in the Mass Gap
\item Decay modes for bound $e^{-}e^{+}$ metastable states
\item Cross section as a function of energy
\item Free $e^{-}e^{+}$ energies in the Ion Center of Mass Frame
\item Free $e^{-}e^{+}$ energies transformed to the Lab Frame
\item Comparison between Experimental Peaks and bound $e^{-}e^{+}$ metastable transitions
\item Explanation of Tables 1 and 2
\end{enumerate}
Section V is a summary which is followed by two appendices deriving Ehrenfest equations and one discussing induced metastable state decays.

\section{Describing the two alternatives}
In this section the motivation for the $D2$ alternative and its distinction from $D1$ is delineated.

\subsection{Free Particle Equations}

Dirac\cite{Diracbook} sought matrices $\bm{\alpha}$ and $\beta$ so that
the free particle Hamiltonian could be written ($c=1$ and $\hbar=1)$ for
momentum $\bm{p}$ and mass $m$ as%
\begin{equation}
H_{0}=\bm{\alpha}\cdot\bm{p}+\beta  m .
\end{equation}

The eigenstates of this Hamiltonian were of two types: positive energy
$u(\bm{p})$ and negative energy $v(\bm{p})$ and each of these
satisfied the equations%
\begin{equation}
(\bm{\alpha}\cdot \bm{p}+\beta \mit m)\mit u(\bm{p})=\mit E_{\bm{p}}\mit u(\bm{p}),
\end{equation}%
\begin{equation}
(\bm{\alpha}\cdot \bm{p}+\beta \mit m)\mit v(\bm{p})=-\mit E_{\bm{p}}\mit v(\bm{p}),
\end{equation}
and where
\begin{equation}
\mit E_{\bm{p}}=\sqrt{\bm{p^{2}} + \mit m^{2}}.
\end{equation}
One of the most important ideas that Dirac discovered was the charge
conjugation symmetry that connected these two eigenvectors to each other and
enabled an understanding of the positron. \ To briefly review this symmetry,
one takes the negative energy eigenvalue equation and changes $\bm{p}%
\rightarrow-\bm{p}$ and takes the complex conjugate of the
equation\cite{Merzbacher}. \ This yields%
\begin{equation}
(-\bm{\alpha}^{\ast}\cdot \bm{p}+\beta^{\ast} \mit m)\mit v^{\ast}(-\bm{p})
=-\mit E_{\bm{p}}\mit v^{\ast}(-\bm{p}),
\end{equation}
where use of $E_{\bm{p}}=E_{-\bm{p}}$ has been made and we now
seek a matrix $C$ that has the following properties%
\begin{equation}
\bm{C}\bm{\alpha}^{\ast}\mit \bm{C}^{-1}=\bm{\alpha}%
\end{equation}
and
\begin{equation}
\bm{C}\beta^{\ast}\bm{C}^{-1}=-\beta.
\end{equation}

Substituting these into Eqn. 5 yields%
\begin{equation}
(-\bm{\alpha}\cdot \bm{ p}-\beta \mit m)\bm{C}v^{\ast}(-\bm{p}%
)=-\mit E_{\bm{p}}\bm{C}v^{\ast}(-\bm{p}),
\end{equation}
which becomes the positive energy eigenvalue equation after factoring out
$-1.$ \ This means that within a phase factor we have the following connection
between the two different energy-signed eigenstates%
\begin{equation}
u(\bm{p})=\mit \bm{C}v^{\ast}(-\bm{p}).
\end{equation}

Thus, the energy spectrum is symmetric about zero. \ For every negative energy
eigenvector $v(\bm{p})$ at energy $-E_{\bm{p}}$, there is a
corresponding positive energy eigenvector at $+E_{\bm{p}}$ which is
given by the charge conjugation transformation. \ In the next section the
extension of these free particle results to include potential energies
following Dirac alternative $D1$ and the alternative path $D2$ will be delineated.

\subsection{Incorporating Potential Energy}

Given the symmetries of the non-interacting Dirac equation, the next question
is how to add in a potential energy. \ Intuitively, for the positive energy eigenstates the
inclusion of a small potential $V$ should have the form%
\begin{equation}
E=E_{\bm{p}}+V,
\end{equation}
and should give the correct non-relativistic limit as the momentum goes to
zero,%
\begin{equation}
E=\sqrt{\bm{p}^{2}+\mit m^{2}}+V\approx m+\frac{\bm{p}^{2}}{2m}+V.
\end{equation}

For the negative energy eigenstates, there is no clear criterion for choosing
the sign of the potential energy to be included. \ There are two possibilities
that could make sense: \

Alternative $D1$: (Dirac's choice) is simply to add the potential to the
negative energy eigenvalue,%
\begin{equation}
E=-E_{\bm{p}}+V=-\sqrt{\bm{p}^{2}+\mit m^{2}}+V\approx-m-\frac
{\bm{p}^{2}}{2m}+V.
\end{equation}

Alternative $D2$: (The path not taken) would add the potential with a negative
sign%
\begin{equation}
E=-E_{\bm{p}}-V=-\sqrt{\bm{p}^{2}+\mit m^{2}}-V\approx-m-\frac
{\bm{p}^{2}}{2m}-V.
\end{equation}

At this moment there is actually little to distinguish or justify either
alternative. \ The first alternative $D1$ is the simplest, applying the same rule
to both energy signed eigenvalues. \ The $D2$ alternative restores the
symmetry about zero energy that the $D1$ alternative seems to destroy. \

The $D1$ alternative has already been shown to be relativistically invariant
under Lorentz transformations and leads to an immediate proof that the
interacting Dirac equation is covariant\cite{BandD}. \

Is it possible that the $D2$ alternative can be made in a Lorentz invariant
fashion that will allow a proof that the resulting Dirac equation is
covariant? \ The first step in such a demonstration requires the review of the
Casimir projection operators $B_{+}(\bm{p})$ and $B_{-}(\bm{p}%
)$\cite{Merzbacher}
\begin{equation}
B_{+}(\bm{p})\mit u(\bm{p})=\mit u(\bm{p}),\;\mit B_{+}%
(\bm{p})\mit v(\bm{p})= 0
\end{equation}
and%
\begin{equation}
B_{-}(\bm{p})\mit v(\bm{p})=\mit v(\bm{p}),\;\mit B_{-}%
(\bm{p})\mit u(\bm{p})= 0.
\end{equation}

The sign of the energy is a Lorentz invariant. \ The positive and negative
energy states do not mix under Lorentz transformations\cite{BandD}. \ Using
these Casimir operators an operator that yields the Lorentz invariant, the
sign of the energy, can be written as%
\begin{equation}
sgn(\mit E(\bm{p}))=\mit B_{+}(\bm{p})-\mit B_{-}(\bm{p}).
\end{equation}

So, a Lorentz invariant way to write the $D2$ alternative is to substitute
$sgn(E(\bm{p)})\mit V$ for the potential energy in the Dirac equation.
\ Then Dirac's equation with a potential energy would have the form%
\begin{equation}
(\bm{\alpha}\cdot \bm{p}+\beta \mit m+sgn(E)V)\psi(\bm{p})=\mit E\psi
(\bm{p}),
\end{equation}
which results in the positive eigenvalue equation being written as%
\begin{equation}
(\bm{\alpha}\cdot \bm{p}+\beta \mit m+V)u(\bm{p})=\mit Eu(\bm{p}),
\end{equation}
and the negative energy eigenvalue equation being written as%
\begin{equation}
(\bm{\alpha}\cdot \bm{p}+\beta \mit m-V)v(\bm{p})=-\mit Ev(\bm{p}).
\end{equation}

If we again apply the charge conjugation transformation, we see that the
energy spectrum in the presence of a potential continues to have the symmetry
that was present for the free particle solutions.  Charge conjugation transformation invariance is restored.\

If the more general case of coupling to electromagnetic fields is considered,
the minimal coupling interaction for this $D2$ alternative in the Dirac
equation should be replaced by
\begin{equation}
p^{\mu}\rightarrow p^{\mu}-sgn(E)eA^{\mu},
\end{equation}
where the charge of the electron is $e$ ($e<0$) and we attempt to follow the sign conventions of Bjorkan and Drell\cite{BandD}.
Because the $sgn(E)$ is a Lorentz invariant, there is no change in the vector
nature of this transformation and the usual proof of covariance carries
through automatically. \

\section{ Comparisons between $D1$ and $D2$}

In this section several properties of both the $D1$ and $D2$ alternatives will
be reviewed. \ Several rather striking differences will be found and the
reader is urged to keep in mind that this comparison will in some cases be
jarring to our intuitions, schooled as we are in the $D1$ alternative. \

\subsection{Charge Conjugation Invariance with potentials in $D2$}

The eigenvalue equation for the Dirac Hamiltonian in the $D2$ alternative
will now have the form%
\begin{equation}
(\bm{\alpha}\cdot(\bm{p}-\mit sgn(E)e\bm{A})+\beta \mit m+sgn(E)e\phi
)\psi(\bm{p})=\mit E\psi(\bm{p}).
\end{equation}
For a positive energy state we would have%
\begin{equation}
(\bm{\alpha}\cdot(\bm{p}-\mit e\bm{A})+\beta \mit m+e\phi)u(\bm{p}%
)=\mit Eu(\bm{p})
\end{equation}
and for a negative energy state
\begin{equation}
(\bm{\alpha}\cdot(\bm{p}+\mit e\bm{A})+\beta \mit m-e\phi)v(\bm{p}%
)=-\mit Ev(\bm{p}),
\end{equation}
where $u(\bm{p})$ and $v(\bm{p})$ are used to represent the
positive and negative energy solutions in the presence of the vector potential
$A^{\mu}.$  Notice that the positive energy equation in $D2$ is exactly the same as in $D1$, so, for example, the
positive energy hydrogenic solutions with the Coulomb potential will be the same.
It may be likely that a reader might confuse alternative $D2$ with an earlier alternative studied by M\"{u}ller, Rafelski,
and Soff\cite{MullerRafelskiSoff}.  In this M\"{u}ller alternative the Coulomb potential was assumed
to be multiplied by the $\beta$ matrix.  For a hydrogenic solution with a Coulomb potential this altered Dirac
equation can be solved analytically, but the atomic spectra does not agree with experiment.  The alternative $D2$
is completely different from this earlier model. In $D2$ the positive energy hydrogen is completely
unchanged and thus all of the atomic states that agree with experiment are recovered.

If we now take the negative energy equation and change $p\rightarrow-p$ and
take the complex conjugate we obtain%
\begin{equation}
\bm{\alpha}^{\ast}\cdot(-\bm{p}+\mit e\bm{A})+\beta^{\ast}%
\mit m-e\phi)v^{\ast}(-\bm{p})=-\mit Ev^{\ast}(-\bm{p}).
\end{equation}
Looking for a matrix $C$ exactly as before, we obtain%
\begin{equation}
(-\bm{\alpha}\cdot(\bm{p}-\mit e\bm{A})-\beta \mit m-e\phi)Cv^{\ast}(-\bm{p})%
=-\mit ECv^{\ast}(-\bm{p}),
\end{equation}
which becomes the positive energy equation after factoring out the $-1$ and we
have the charge conjugation relationship as before \
\begin{equation}
u(\bm{p})=\mit \bm{C}v^{\ast}(-\bm{p}).
\end{equation}

Thus, this $D2$ alternative restores the invariance of the Dirac equation
under charge conjugation transformations even in the presence of a vector potential.

\subsection{Ehrenfest Equations in $D2$}
As discussed in the introduction, the Ehrenfest equations in $D1$ for the classical Lorenz force depended on the sign of the
energy of the wave function evaluated in the expectation values.  A similar situation
transpires for a spin in a constant magnetic field.

The first of these is the calculation of the change in the momentum in the
presence of a vector potential and a scalar potential. \ In the Dirac $D1$
alternative the derived equation contains the sign of the energy of the
eigenstates used to calculate the expectation values\cite{Merzbacher}. \ The
derivation is outlined in Appendix A%
\begin{equation}
sgn(E)\frac{d}{dt}(\langle m\bm{v\rangle})=-e(\bm{E}%
+\bm{v}\times\bm{B}).
\end{equation}
When the minimal coupling for alternative $D2$ is included, a second factor of
$sgn(E)$ multiplies the electric and magnetic fields on the right hand side
and cancels out the factor on the left hand side. \ This recovers the classical
Ehrenfest formula for the momentum%
\begin{equation}
\frac{d}{dt}(\langle m\bm{v\rangle})=-e(\bm{E}+\bm{v}%
\times\bm{B}).
\end{equation}

In Appendix B a similar calculation for the spin dynamics of an electron in a
magnetic field and factor of $sgn(E)$ for the eigenstates forming the
expectation value is also found in the same way as for the Lorentz force law
in alternative $D1$. \ In alternative $D2$ a second factor of $sgn(E)$ is
found on the right hand side and the appropriate Ehrenfest equation is recovered.
So, $D2$ corrects the inconsistencies of the Ehrenfest equations in $D1$.

\subsection{Choice of one electron basis states in $D2$}

In both alternatives the matrix elements of the Hamiltonian can be determined relative to
any chosen basis set.
\ The usual approach is to begin with positive energy free electron plane waves. \ An equally
valid basis set could be derived from eigenstates of hydrogenic atomic
states or the eigenstates of any potential. The hydrogenic set of states would be more appropriate for situations
involving a nucleus at the origin as found in heavy ion scattering and other atomic problems.
In the mass gap this hydrogenic basis will describe the bound states quite well.  For energies outside of the mass gap
these eigenstates, though derived from a nucleus at the origin could equally well describe scattering waves about the origin as well
as other potentials.
\ At large energies and large distance from the nucleus the hydrogenic
states are essentially coulomb scattering states which become approximately plane wave-like
 at very large distances. \ These basis states reflect the dominant
influence of the charged nucleus. \ If there are other charges, as in
scattering from other atoms, it may be desirable to distinguish between the
internal potential which gives rise to the hydrogenic states and the external
potential that could describe scattering in this basis based on the central
nucleus. \ In both of the alternatives, the scattering states at high
energies and large distances will behave very much like the free electron bases.
However, in $D2$ the atomic hydrogenic
states will exhibit bound states in both the top and bottom halves of the mass
gap because the energy specrum must be symmetric about zero.  Electron and positron
scattering results should be able to be described by any similar bases.  The
distinguishing characteristic will be the bound states in the gap.

\subsection{Bound States in the Mass Gap in $D1$ and $D2$}

When bound states in the mass gap are considered, there are very real
differences between the two alternatives. \ In alternative $D1$ , there are
only bound states in the positive half of the mass gap and no states in the
lower half of the mass gap. \ In alternative $D2$ the positive energy bound
states in the upper part of the mass gap are mapped by the charge conjugation
transformation into the lower half of the gap. \ The existence of these bound
states in the lower half of the mass gap gives rise to a new set of excitations
that do not exist in the $D1$ alternative. \ These negative energy bound
states in the lower part of the gap must also be filled in the Dirac vacuum.
\ Excitations out of these states will leave behind bound holes in the vacuum that
are localized near the nucleus. \ The existence of these states should give rise
to excitations that were not expected in alternative $D1.$ \ It should be
possible to distinguish between the two alternative by examining the
properties of these hole states. \ That is the goal of a later section in this
paper which examines data from heavy ion scattering experiments mentioned
above. $D2$ displays a different energy spectrum in the mass gap from $D1$.\

\subsection{Piece-wise Constant Potentials in Alternative $D1$ and $D2$}

The examination of constant potentials is important because any potential can
be approximated mathematically by piecewise constant potentials. \ In the following
discussion the calculation will be carried out in one dimension for spinless
particles since it simplifies the mathematics. \ This reduces the matrices to
$2\times2$'s and spinors to $1\times2$'s.

Consider first a positive constant potential $V_{0}$ in a certain region. \ In
this region the wave function
\begin{equation}
\psi(z)=ae^{ikz}
\biggl\lbrack
\begin{array}{c}
u_{1}\nonumber\\
u_{2}\nonumber
\end{array}
\biggr\rbrack
\end{equation}
will carry a current in the positive $z$ direction. \ Now consider a
comparison of this wave function and its energy spectrum $E$ in both the $D1$
and $D2$ alternatives.

In alternative $D1$, the Dirac equation yields the following secular matrix
equation%
\begin{equation}%
\biggl\lbrack
\begin{array}{cc}
V_{0}-E+m & k\\
k   & V_{0}-E-m
\end{array}
\biggr\rbrack
\times
\biggl\lbrack
\begin{array}{c}
u_{1}\nonumber\\
u_{2}\nonumber
\end{array}
\biggr\rbrack
=0.
\end{equation}

It is easy to show that the energy eigenvalues are%
\begin{equation}
E_{k}^{\pm}=V_{0}\pm\sqrt{m^{2}+k^{2}},
\end{equation}
and the wave functions are%
\begin{equation}
\psi_{k}^{+}(z)=ae^{ikz}%
\biggl\lbrack
\begin{array}{c}
1 \nonumber \\
\frac{k}{m+\sqrt{m^{2}+k^{2}}}\nonumber
\end{array}
\biggr\rbrack
,
\end{equation}
and%
\begin{equation}
\psi_{k}^{-}(z)=ae^{ikz}%
\biggl\lbrack
\begin{array}{c}
-k\over(m+\sqrt{m^{2}+k^{2}})\nonumber\\
1 \nonumber
\end{array}
\biggr\rbrack
.
\end{equation}

The defining signs of the energies are relative to $V_{0}$ and the actual sign
of the energy will depend on the size of $V_{0}.$

The "positive" energy states are found for $E>V_{0}+m$ and the "negative"
energy states are found for $E<V_{0}-m$ and there are no states found in the
mass gap $V_{0}-m<E<V_{0}+m$ which is centered on $V_{0}.$ \ As $V_{0}$
becomes more positive, the "negative" (relative to $V_{0})$ states are pulled
up in energy and can become degenerate with positive energy states in a nearby
region, leading to Klein paradox effects. \ These effects were shown in Fig.
1 for the energy range $(m,V_{0}-m)$ if $V_{0}$ is large enough. \
Specifically, for a potential step of height $V_{0} $
and extending infinitely far in one direction, the transmission coefficient
$T$ can be calculated analytically.  If the potential step height $V_{0}$ is
less than the width of the mass gap,
$V_{0}<2m$, (in this section $\hbar=1$ and $c=1$), the transmission coefficient $T$
is zero for incident particle
energies $E$ in the range $m<E<V_{0}+m$ and becomes non-zero for $E>V_{0}+m.$
\ $T$ approaches $1$ as the incident energy becomes large. \ These results are
very classical. \

If the potential step is larger so that $V_{0}>2m,$ the transmission
coefficient develops a non-zero value between $m$ and $V_{0}-m.$ \ It is as if
the barrier has become partially transparent to particles in the range
$m<E<V_{0}-m.$ \ Mathematically, this effect is due to the negative energy
states (relative to $V_{0})$ being pulled up into degeneracy with the positive
energy states of the incident particles. \ An example of the transmission
coefficient for this case is shown in Fig. 1, for the value of $V_{0}/m=3.$ \ This situation
of Klein tunneling presents a serious challenge for any theory of low energy particles
because they would appear to tunnel into high and infinitely wide barriers with very little
energy. \ This would seem to contradict widely observed classical phenomena.

\begin{figure}
\includegraphics[width=\columnwidth]{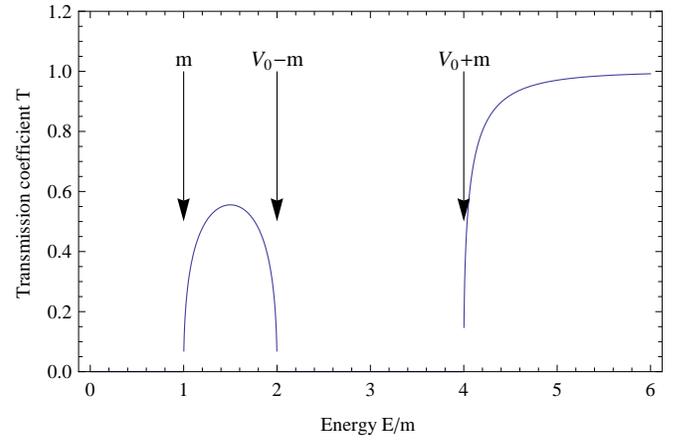}
\caption{The one dimensional transmission Coefficient $T$ (calculated in alternative D1 )
for a potential step of height $V_{0}/m=3$ for $0<x<\infty$ plotted as a function of the incident particle
energy $E$.  The small peak between $m<E<V_{0}-m$ is the Klein tunneling.}
\label{fig:1}
\end{figure}

The situation for the $D2$ alternative is quite different. \ The secular
matrix equation becomes%
\begin{equation}%
\biggl\lbrack
\begin{array}{cc}
sgn(E)V_{0}-E+m & k\nonumber\\
k & sgn(E)V_{0}-E-m\nonumber
\end{array}
\biggr\rbrack
\times
\biggl\lbrack
\begin{array}{c}
u_{1}\nonumber\\
u_{2}\nonumber%
\end{array}
\biggr\rbrack
=0
\end{equation}
and the eigenvalues satisfy%
\begin{equation}
sgn(E)(|E|-V_{0})=\pm\sqrt{m^{2}+k^{2}}.
\end{equation}
So, for the positive sign the eigenvalue is%
\begin{equation}
E_{k}^{+}=V_{0}+\sqrt{m^{2}+k^{2}},
\end{equation}
while for the negative sign%
\begin{equation}
E_{k}^{-}=-V_{0}-\sqrt{m^{2}+k^{2}}%
\end{equation}
and the corresponding wave functions are unchanged.

In this $D2$ alternative the positive energy states are found for $E>V_{0}+m$
and the negative energy states are found for $E<-(V_{0}+m).$ \ There are no
oscillatory states allowed in the intervening interval $-(V_{0}+m)<E<V_{0}+m.
$ \ Since there are no states in this interval, there is no possible current
in this energy range. \ In contrast to the $D1$ alternative, in the $D2$
alternative increasing the strength of the potential simply widens the
interval in which no states are allowed. \ It is not possible for "negative"
energy states to ever be degenerate with positive energy states in a
neighboring region.  The $D2$ alternative suppresses the Klein tunneling allowed in $D1$.

\subsection{Klein Tunneling in $D1$ and $D2$}

Consider the classic step potential of the Klein paradox ($V=0$ for $z<0$ and
$V=V_{0}$ for $z>0$ )in the $D2$ alternative.  There are no
states for $z>0$ in the energy interval $m<E<V_{0}+m,$ and the transmission
coefficient $T$ must be zero in this interval. \ For energies $E>V_{0}+m$ the
transmission coefficient $T$ is given by%
\begin{equation}
T=\frac{4r}{(1+r)^{2}},
\end{equation}
where%
\begin{equation}
r=\sqrt{\frac{E-(V_{0}+m)}{E-(V_{0}-m)}}\sqrt{\frac{E+m}{E-m}}%
\end{equation}
and $E=\sqrt{m^{2}+k^{2}}$ is the incident energy. \

In the $D2$ alternative there is no Klein paradox. \ Increasing $V_{0}$ simply
expands the interval $(m,V_{0}+m)$ where $T=0.$ \ If Fig. 1 were constructed
in the $D2$ alternative, there would be no small peak between $m$ and
$V_{0}-m$ and $T=0$ in that interval. \ There is no question about the
potential being too strong. \ There is simply no current in the step in the
energy interval $(m,V_{0}+m).$

In this classic step problem it is possible to find evanescent waves whose
energies are less than $V_{0}+m.$ \ These evanescent waves do not carry
currents, so their presence does not change the discussion about $T$ vanishing
in the interval $(m,V_{0}+m).$

In the $D2$ alternative the evanescent wave function for $z>0,$ for example,
will have the following form%
\begin{equation}
\psi(z)=ae^{-\kappa z}%
\times
\biggl\lbrack
\begin{array}{c}
w_{1}\nonumber\\
w_{2}\nonumber%
\end{array}
\biggr\rbrack
\end{equation}
and the eigenvalues are found to be%
\begin{equation}
E_{\kappa}^{+}=V_{0}+\sqrt{m^{2}-\kappa^{2}}%
\end{equation}
and%
\begin{equation}
E_{\kappa}^{-}=-V_{0}-\sqrt{m^{2}-\kappa^{2}}%
\end{equation}
where $0<\kappa<m$ and for the positive energy state $w_{1}=1$ and
$w_{2}=i\kappa/(m+\sqrt{m^{2}-\kappa^{2}}).$ \ No states are allowed in the
interval $(-V_{0},V_{0}).$ \ These evanescent states would enable tunneling in
the energy interval $(V_{0},V_{0}+m),$ but not in the interval $(-V_{0}%
,V_{0}).$

It should be noted again that the energy spectrum of these states are symmetric
about the zero energy.  $D2$ does not allow the Klein tunneling effects which are in $D1$.

\subsection{Attractive Potentials in $D1$ and $D2$}

In $D1$ a negative square well can have any depth.  It is possible to have bound states in the gap of
such a potential at any depth in the mass gap.  It was the question of what happens when the potential is deeper than the mass gap
and bound states of this potential were to merge with the continuum of negative states below the mass gap
that motivated the unsuccessful search for unstable vacua by Greiner and colleagues.

In $D2$ the presence of $sgn(E)$ gives rise to a surprising difference from $D1$.  \ Consider a negative potential $V(x)$ (for the moment
restrict the potential to be one dimensional). \ Consider first a positive
energy eigenstate of this system. \ This means that the $sgn(E)$ factor multiplying the
potential is positive. \ For a negative potential the eigenvalues that are
lower than $m$ will be found in the mass gap $(-m,m).$ \ How deep can the
potential be and still give rise to a positive energy eigenvalue? \ Clearly if
the depth of the potential is less than $m$ the bound states will be above
zero and will clearly have positive energies. \ However, if a potential
extends from $+m$ down to some negative energy below zero, it is possible that
the potential could have a bound state that is below zero, that is negative,
but we have been considering only positive energies and this would lead to a
contradiction. \ Thus, this \it{reductio ad absurdum} \rm argument would imply that in alternative $D2$, a
negative potential cannot be deeper than $-m$ or more precisely, cannot have bound states that are below zero energy.
Because of the charge conjugation symmetry in $D2$, this would mean that the energy spectrum remains symmetric about zero and the eigenstates
and eigenvalues are mapped from positive energy onto negative energies.  There cannot be any crossing through zero.  \

From this perspective it is notable that the positive energy hydrogenic bound states of the
Dirac equation are clearly in the interval $(0,m)$ and that the lowest
possible state for a nuclear charge $Z$ is zero as $Z\rightarrow 1/\alpha_{0},$ where
$\alpha_{0}$ is the fine structure constant.

In this $D2$ alternative a plot of the potential energies for both positive
and negative energy states will have a reflection symmetry about zero energy and
never cross over the zero energy. \ No attractive potential for negative energy states can extend
above the zero energy level. \ There does not seem to be any such restriction
on repulsive potentials since any eigenvalues would be larger than $m.$
\ There does not seem to be any obvious restriction on the height of barriers.

$D2$ seems to imply that no positive energy state can cross the zero energy line as is allowed in $D1$.  This
property explains why the search for an unstable vacuum was not successful.
The ultimate question is whether nature actually exhibits distinctive behavior
that confirms the $D2$ alternative as the correct alternative.

\subsection{Hole Theory in $D1$ and $D2$}

Let us now consider Dirac's argument for positrons in the hole theory in both
$D1$ and $D2.$ \ First, it must be re-iterated that in the free electron case
there is explicitly no charge in the Hamiltonian. \ Charge only enters the
description in the presence of a vector potential. \ Charge
enters into the hole theory in an independent fashion through an argument for the change in charge of the
vacuum in the presence of a hole. \

In the absence of any potential both $D1$ and $D2$ are the same. \ The Dirac
vacuum with all of the negative energy states filled is necessary to avoid
transitions to unbounded negative energies. \ In both alternatives a hole in
the free electron vacuum at an energy of $-E_{p,}$ momentum $-p$ , and spin
$-s$ is clearly interpreted by the change in the energy of the many electron
state as a positive energy state $E_{p,}$ with momentum $p$ and spin $s.$
\ The charge associated with a hole is imputed by observing the change in the
charge of the vacuum plus hole in comparison with the vacuum. \ The charge in
not carried by the wave function. \ The charge is the opposite of whatever the
charge was of the original electrons whose wave functions were solutions of
the free electron Dirac equation. \

$D2$ describes positron wave functions using charge conjugation in the same way as does $D1$.
However, the positron wave function in $D2$ is identical with some positive energy electron
wave function in contrast to $D1$.  The charge is not carried in the wave function, but is ascribed by
the change in the vacuum as in $D1$.

In $D1$ there is a long history of ascribing of the charge of the vector potential in the Dirac
equation under the charge conjugation transformation as evidence that the positron wave function represents
an opposite charge from the electron.  From the perspective of $D2$ this sign change is simply the breakdown
of charge conjugation invariance in $D1$.  In $D2$ the positron wave function must necessarily be identical
to a positive energy electron state just as is true for the free electrons.  The positron charge is not carried inherently in
the wave function and must be imputed separately.

\subsection{Current Density and Zitterbewegung  in $D1$ }

A central argument for the $D1$ alternative has been the discussion\cite{BandD} of the
probability current density for free electrons. \ The standard derivation for
the continuity equation of the probability density yields for the probability
current density%
\begin{equation}
J_{i}=c\alpha_{i}%
\end{equation}
where $c$ is the velocity of light and $\alpha_{i}$ is one of the
Dirac matrices. \

This result has long been regarded as peculiar since in the classical limit we
expect the current density to be carried by the momentum $p_{i}/m.$ \ A long
standing puzzle related to this peculiarity is the fact that in the absence of
a potential the momentum is a constant of the motion. \ However, this
probability current density does not commute with the Hamiltonian and is not a
constant of the motion in the absence of interaction potentials. This is the third Ehrenfest
equation in $D1$ which does not obey the classical limit.

In the $D1$ alternative using a wave packet constructed out of positive energy states%
\begin{equation}
\psi_{(+)}(x)=\int\frac{d^{3}p}{(2\pi\hbar)^{3/2}}\sqrt{\frac{m}{E_{p}}}%
\sum_{s}b(p,s)u(p,s)e^{-ip\cdot x},
\end{equation}
it is straight forward to evaluate the expectation value of the probability
current density and show that%
\begin{equation}
\langle J_{i}\rangle_{+}=\langle c\alpha_{i}\rangle_{+}=\int
\frac{d^{3}p}{(2\pi\hbar)^{3/2}}\frac{c^{2}p_{i}}{E_{p}}|b(p,s)|^{2}%
=\langle\frac{c^{2}p_{i}}{E_{p}}\rangle_{+}%
\end{equation}
where the subscript $+$ on the averages indicates only the positive energy states are used.

At this point a crucial argument has been made in the $D1$ alternative. \ It
is noted that the eigenvalues of the $c\alpha_{i}$ matrices are
$\pm c$, corresponding to positive and negative energies, and if the expectation value of the probability current density is to
be calculated using the eigenvectors of the $\alpha_{i}$ matrices
it will require the inclusion of negative energy states in order to obtain a velocity
associated with the
probability current density less than the velocity of light. \

When the current density is evaluated using such a linear combination of
states,%
\begin{eqnarray}
\psi(x)=\int\frac{d^{3}p}{(2\pi\hbar)^{3/2}}\sqrt{\frac{m}{E_{p}}}\sum
_{s}(&&b(p,s)u(p,s)e^{-ip\cdot x}  \nonumber \\
&&+ d^{\ast}(p,s)v(p,s)e^{ip\cdot x}),
\end{eqnarray}
the result\cite{BandD} is a weighting of the momentum by the amplitudes $|b(p,s)|^{2}$ and
$|d(p,s)|^{2},$ and also the famous zitterbewegung terms that exhibit
frequencies on the order of the rest mass $\pm 2 mc^{2}.$ \ These high frequency
terms have been puzzling and have led to a number of intuitive arguments
about confinement and the excitation of electron-positron pairs in the
presence of static potentials.

The amplitude of the zitterbewegung in the expectation value of the
probability current density is proportional to the amplitude $d^{\ast}(p,s)$
of the negative energy states in the wave function. \ The next series of
$D1$ arguments are guided by an example wave function. \ A typical wave
function\cite{BandD} at zero time is
\begin{equation}
\psi(x)=(\pi d^{2})^{-3/4}e^{-\frac{x^{2}}{2d^{2}}}w_{1}(0)
\end{equation}
where $d$ is the "confinement width" of the wave function and $w_{1}(0)$ is
the spin up spinor for an electron with zero momentum and positive rest energy.

The amplitude of a positive energy plane wave in this state is%
\begin{equation}
b(p,s)=A(\frac{d^{2}}{\pi\hbar^{2}})^{3/4}%
e^{-\frac{p^{2}d^{2}}{2\hbar^{2}}},
\end{equation}
and the amplitude for a negative energy state is%
\begin{equation}
d^{\ast}(p,s)=A(\frac{d^{2}}{\pi\hbar^{2}}%
)^{3/4}e^{-\frac{p^{2}d^{2}}{2\hbar^{2}}}\frac{pc}{E_{p}+mc^{2}},
\end{equation}
where $A$ is a normalization factor.

The relative fraction of the negative energy states in this wave function is
small unless $pc\approx mc^{2}$ or greater. \ At the same time the confinement
of the wave function to a region of size $d$ implies the momentum must
be $p\leq\hbar/d,$ or that the confinement is comparable to the Compton
wavelength%
\begin{equation}
d\approx\frac{\hbar}{mc}.
\end{equation}

From these arguments in $D1$ there have evolved two conclusions. \ If the
confinement of a wavefunction is on \ the order of the Compton wavelength or
smaller, negative energy states will be significantly probable. \ Because of
the connection between negative energy states and positrons this would mean
that static, confining potentials should generate electron-positron pairs under the conditions of
close confinement. \

The origin of these arguments in $D1$ is the fact that the probability current
density is not proportional to the momentum operator and to get a speed
associated with the probability current density to be less than the velocity
of light, negative energy states must be mixed into any wave function. \ The
relevant question, besides the question of whether all of the inconsistencies
of $D1$ are acceptable, is whether the probability current density is what
transports the charge density. \ This question is resolved in $D2$ in the next
section. \

\ These ideas are deeply engrained in those of us who have learned
quantum mechanics in alternative $D1.$ \ In particular, we have been forced
into these outcomes by the form of the probability current density. \ Let us
now examine the same picture from the point of view of alternative $D2.$ \

\subsection{Charge Current Density in $D2$}

First, it must be noted that both alternatives $D1$ and $D2$ will have the same probability
current density for free electrons. \ For free electrons there is no
distinction between the two alternatives. \ However, there is also explicitly no charge in
the free electron Hamiltonian. To examine the charge current density it is necessary to examine the Hamiltonian
that couples the electrons to a vector potential.

The $D2$ alternative differs primarily in the treatment of the electron
equations in the presence of a vector potential $A_{\mu}=(\Phi,A_{i}).$ For a Hamiltonian that couples vector potentials to the free
electrons, the charge current density $J_{i}^{Q}$ is most easily derived by
\begin{equation}
J_{i}^{Q}=-\frac{\partial H}{\partial A_{i}}.
\end{equation}

Since the Hamiltonian in alternative $D2$ is
\begin{equation}
H=c\bm{\alpha}\cdot\bm{p}+\beta mc^{2}- e\> sgn(E)\bm{\alpha
}\cdot\bm{A}+ e\> sgn(E)\Phi,
\end{equation}
the charge current density is%
\begin{equation}
J_{i}^{Q}=esgn(E)\alpha_{i}={\frac{1}{2}} e\>(sgn(E)\alpha_{i}+\alpha_{i}sgn(E)),
\end{equation}
where in the second equality the charge current density operator has been symmetrized
as needed for the free electron Hamiltonian.
For free electrons with momentum $\bm{p}$ , we have that $sgn(E)$ can be represented by
a projection operator%
\begin{equation}
sgn(E)=\frac{c\bm{\alpha}\cdot\bm{p}+\beta mc^{2}}{E_{p}}.
\end{equation}
Substituting the expression for the free electron projection operators into the expression for $J_{i}^{Q}$ we obtain
\begin{equation}
J_{i}^{Q}=e(1/2)( c\bm{\alpha}\cdot\bm{p}\alpha_{i}+c\alpha_{i}\bm{\alpha}\cdot\bm{p} + mc\lbrace{\beta,\alpha_{i}}\rbrace )
\end{equation}
and using the anti-commutation relations for the $\alpha_{i}$ and $\beta$ matrices
\begin{equation}
\lbrace\alpha_{i},\alpha_{j}\rbrace=2\delta_{i,j}
\end{equation}
\begin{equation}
\lbrace\alpha_{i},\beta\rbrace=0
\end{equation}
 we obtain the current density operator proportional to a component of the momentum operator%
\begin{equation}
J_{i}^{Q}=I_{4}ec\frac{p_{i}}{E_{p}},
\end{equation}
the matrix $I_{4}$ is the $4x4$ identity matrix and $p_{i}$ is the i-th
component of the momentum operator. \

Thus, in alternative $D2$ the charge current density is in excellent accord
with classical expectations. \ It is proportional to the momentum, is a constant of the
motion in the absence of a potential, thus satisfying the Ehrenfest equations
which were violated in alternative $D1.$ \ Because the charge current density is proportional
to the momentum there is no zitterbewegung!  There is no paradox as in $D1$
where the probability current density operator has eigenvalues which force the
inclusion of negative energy states in wave packets.  \ Expectations of the current density can be determined
completely by positive energy wave packets as above. There is no necessity for the inclusion
of negative energy states in wave function packets as was argued in $D1$.  .
\ In alternative $D2$ the
negative energy states are derived from the positive energy states by the
charge conjugation transformation and only the
positive energy states are necessary for completeness. \

The arguments that confinement by static potentials necessarily induces
particle anti-particle states in so far as it was based on arguments of
eigenvalues of $c\alpha$ no longer has any imperative in alternative $D2.$  In $D2$ the charge
current density, which is what describes the movement of the charge, is proportional to the momentum and there is
no zitterbewegung or generation of positrons by confinement in static potentials as argued in $D1$.

\subsection{Completeness in $D2$}

It is often argued in $D1$ that positive energy states are not complete.  This argument relies completely on the
apparent need in $D1$ for negative energy eigenvalues of the $\alpha$ matrices. As has been discussed above
there is no need for this argument when considering the charge current density.  In $D2$ the standard
mathematical proofs based on the Fourier Theorem\cite{MorseandFeshbach} can be used to prove that the positive
energy states are complete in the Hilbert space.  Since the negative energy states are connected to the positive
energy states by the charge conjugation transformation, they are not linearly independent of the positive
energy states.  This means that the positive energy states are complete and are the only set of states needed.
The inferences in $D1$ based on the probability current density and the eigenfunctions of the $\alpha$ matrix
are irrelevant to the charge current density and thus the movement of charge in $D2$.  The arguments on confining
potentials generating $e^{-}e^{+}$ pairs does not arise in $D2$.
In $D2$ the positive energy states are complete in contrast to the conclusions in $D1$.

\subsection{Transformations and Feynman/Stuckleberg Theory in $D1$ and $D2$}

In both of the alternatives $D1$ and $D2$ there are three transformations that
can be formulated in the free electron basis. \ These transformations are well
defined in several references\cite{BandD,Merzbacher,Messiah}. \ They are: $\bm{C}$ charge conjugation,
$\bm{\tau}$ time reversal, $\bm{P}$ parity transformations. \ If we write the
Hamiltonian in two parts:%
\begin{equation}
H_{0}=\bm{\alpha}\cdot \bm{p}+\beta m
\end{equation}
and%
\begin{equation}
H_{1}=-e\bm{\alpha}\cdot \bm{A}+e\Phi,
\end{equation}
it is easy to show the following transformations.%
\begin{equation}
\bm{C}H_{0}(\bm{p})\bm{C}^{-1}=-H_{0}(-\bm{p}),%
\end{equation}%

\begin{equation}
\bm{P}H_{0}(\bm{p})\bm{P}^{-1}=H_{0}(-\bm{p}),%
\end{equation}%
\begin{equation}
\bm{\tau} H_{0}(\bm{p})\bm{\tau}^{-1}=-H_{0}(-\bm{p}).%
\end{equation}
The first of these shows that the free electron Hamiltonian has a
reflection symmetry about zero energy and is charge conjugation invariant.
\ \ By exactly the same process for free electron states, we can show%

\begin{equation}
\bm{C}sgnE(\bm{p})\bm{C}^{-1}=-sgnE(-\bm{p}),%
\end{equation}%
\begin{equation}
\bm{P}sgnE(\bm{p})\bm{P}^{-1}=sgnE(-\bm{p}),%
\end{equation}%
\begin{equation}
\bm{\tau} sgnE(\bm{p})\bm{\tau}^{-1}=-sgnE(-\bm{p}).%
\end{equation}

If we apply these transformations to the interaction part of the Hamiltonian,
we obtain%

\begin{equation}
\bm{C}H_{1}\bm{C}^{-1}=H_{1},%
\end{equation}%
\begin{equation}
\bm{P}H_{1}\bm{P}^{-1}=H_{1},%
\end{equation}%
\begin{equation}
\bm{\tau} H_{1}\bm{\tau}^{-1}=H_{1}.%
\end{equation}

In alternative $D2,$ we have charge conjugation invariance and have the
following transformations.%

\begin{equation}
\bm{C}(H_{0}(\bm{p})+sgnE(\bm{p})H_{1})\bm{C}^{-1}=-(H_{0}
(-\bm{p})+sgnE(-\bm{p})H_{1},%
\end{equation}%
\begin{equation}
\bm{\tau}(H_{0}(p)+sgnE(\bm{p})H_{1})\bm{\tau}^{-1}=-(H_{0}(-\bm{p})+sgnE(-\bm{p})H_{1}).%
\end{equation}
Negative energy wave functions are transformed into positive energy wave
functions and vice versa. \ The charge conjugation wave function $\psi_{c}$of
a negative energy state $\psi_{(-)}$ is given by
\begin{equation}
\psi_{c}=\bm{C}\psi_{(-)}^{\ast}%
\end{equation}
is identical with a positive energy state with the appropriate momentum and
spin and define the positron wavefunction. \ Just as in alternative $D1,$ because the time reversal transformation
gives rise to the same transformation properties as for the charge conjugation, the
time reversed wave function will necessarily be the same (within multiplication by a constant matrix) as the charge
conjugation transformed wave function except for a phase factor. The Wigner time reversed wave function ${\Psi_{PCT}}$ \cite{BandD} is an explicit
representation of this connection between positron wave functions and a matrix multiplied by the wave function of a negative energy electron moving backward
in space time and is valid in $D2$ as well as $D1$.

In alternative $D1$ charge conjugation invariance is broken and we have
\begin{equation}
\bm{C}(H_{0}(\bm{p})+H_{1})\bm{C}^{-1}=-{(H_{0}(-\bm{p})-H}_{1}).%
\end{equation}
In much of the literature about alternative $D1$ this breakdown of the charge
conjugation invariance has been interpreted as evidence for the sign change of
the positron. \ This has persisted even though the positron charge was
determined by the vacuum neutrality condition. \ From the perspective of the
$D2$ alternative, the change of sign is just the breakdown of charge
conjugation invariance. \

In $D2$ it is possible to demonstrate the apparent charge difference by examining the negative energy
wave function and its time reversal transform.
  For a positive energy, the wave functions must obey%
\begin{equation}
i\hbar\frac{\partial\psi_{(+)}}{\partial t}=(\bm{\alpha}\cdot(\bm{p}-e\bm{A})+\beta
m+e\Phi)\psi_{(+)}.
\end{equation}
For a negative energy the wave function must obey an equation that looks like
an oppositely charge particle, except for having a negative energy.%

\begin{equation}
i\hbar\frac{\partial\psi_{(-)}}{\partial t}=(\bm{\alpha}\cdot(\bm{p}+e\bm{A})+\beta
m-e\Phi)\psi_{(-)}.
\end{equation}
The time inversion transformation identifies the transformation of this
negative energy wave function as a positive energy wave function traveling
backward in time.

\subsection{Feynman Diagrams and Perturbations in $D1$ and $D2$}

Because in $D2$ the negative energy states are determined by mapping from the positive energy states
using the charge conjugation transformation and the positron amplitude has been identified with negative energy wave functions moving backward in space time, the application of the whole diagrammatic structure of the Feynman Stuckleberg perturbation structure\cite{BandD} should
be exactly the same in $D1$ and $D2$ when applied on positive energy states.  The negative energy results could then
be obtained by the charge conjugation transformation.

\

\section{Heavy Ion Scattering as evidence for alternative D2}

In this section the experimental energies of the sharply defined $e^{-}e^{+}$ pairs seen in heavy
ion scattering will be compared with predictions of the $D2$ alternative.  The properties of $D2$ delineated in the previous
sections already predict that no evidence of an unstable vacuum would be found because positive and negative energy eigenvalues are
separated by the zero energy.

\subsection{Historical Introduction}

P. A. M. Dirac's \cite{Dirac1} proposal for a \textquotedblright
vacuum\textquotedblright\ in which all of the negative energy electron states
were filled has led to a detailed understanding of free positrons, including
their production and annihilation\cite{BandD}.  However, the extension of his
insights to bound, hydrogenic-like atomic states
of the Coulomb potential has resulted in a strange story of contradictory
experimental and theoretical results. \ Using a widely held version of
alternative $D1$ and vacuum for bound states \cite{GandH}, Greiner and numerous
colleagues\cite{Greineretal1},\cite{Greineretal2} proposed a series of compelling effects to be
seen in the high electric fields of heavy ion scattering experiments. \

These predictions and experiments were the motivation for three international
proceedings: Landstein\cite{Landstein} Maratea\cite{Maratea86} and
Cargese\cite{Cargese90}. \

The 1990 summary lecture\cite{Summary90} by M\"{u}ller, for the last international
meeting \cite{Cargese90}, assembled a table of $e^{-}e^{+}$ pair sum energies
and noted that \textquotedblright the phenomenology of data was so complex
that they do not fit into any simple scheme..\textquotedblright\ and little
hope was offered of finding such a scheme. \ An introductory essay for this
same conference by J. Rafelski\cite{vacuum90} ruminated that striking
experimental observations were needed showing the breakdown of the vacuum in
order to be sure that their understanding not be \textquotedblright
Ptolemean\textquotedblright, that is, making a fundamental mistake in a basic
notion, leading to more and more complex descriptions. \ None of the theories
in this meeting\cite{Cargese90} were able to make sense of the carefully
crafted experimental data.

\subsection{Hydrogenic bound $e^{-}e^{+}$ states in the mass gap}
The charge conjugation invariance in $D2$ gives rise to a very different structure of states in the mass gap.
The positive energy states for the hydrogenic atom are the same in both $D1$ and $D2$.  In $D2$ the charge conjugation invariance maps the positive energy bound states to the lower half of the mass gap.
Shown in Fig. 2 is a representation of the mass energy gap of the Dirac equation in both the $D1$ and $D2$ alternatives.  In each alternative the Dirac vacuum will be formed by assuming all of the negative energy states are occupied by electrons.  In the $D2$ alternative that would include the negative energy bound states in the lower part of the gap.  Of course, in the $D1$ alternative there are no bound states in the bottom of the gap.

\begin{figure}
\includegraphics[width=\columnwidth]{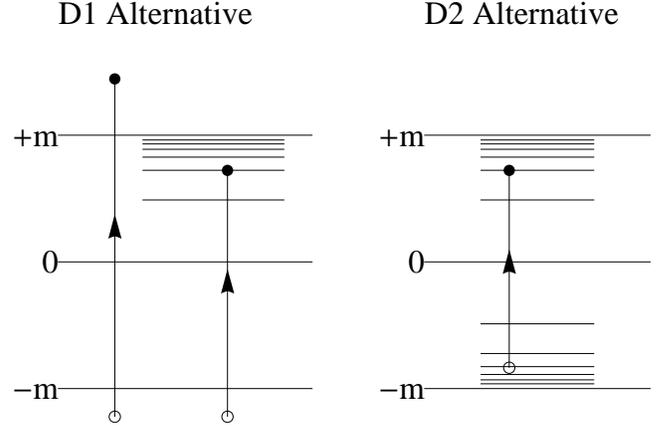}
\caption{Representation of the types of electron positron excitations available in Alternatives $D1$ and $D2$.  The two excitations on the left are available in both $D1$ and $D2$, but the excitation on the right represents a bound $e^{-}e^{+}$ pair that is only possible in alternative $D2$. }
\label{fig:2}
\end{figure}

The excitations on the left side of Figure 2 are present in both $D1$ and $D2$ because they involve holes below the mass gap.  The left most excitation creates a free $e^{-}e^{+}$ pair.  The second excitation creates an excited atomic state and a free positron.

The right side of Fig. 2 is only found in the $D2$ alternative and illustrates an excitation from an occupied negative energy bound state to an (unoccupied) positive energy bound state.  This state will be described as a bound $e^{-}e^{+}$ pair state.  \ These bound $e^{-}e^{+}$\ pairs within an ion
constitute a new excited state of the system. \ The movement of these bound $e^{-}e^{+}$\ pairs is carried by the ions. \ Similarly, vertical
excitations during ion-ion collisions from the filled vacuum states to an empty (ionized) positive energy
state would give rise to a metastable, bound $e^{-}e^{+}$ pair state on one or both of the scattered
ions.\

In the following we will argue that these excitations are responsible for the narrowly defined free $e^{-}e^{+}$ pair energies observed in heavy ion scattering.  For that argument it is first necessary to determine the excitation energies for these bound $e^{-}e^{+}$ pairs.

In the $D2$ hydrogenic approximation the energy levels of the $e^{-}e^{+}$ bound
pairs will be the difference between the positive energy bound hydrogenic
state $E_{+}(n,j)$ and the negative energy bound state $E_{-}(n,j)$ where%
\begin{eqnarray}
E_{\pm}&&(n,j)=  \nonumber \\
&&\pm m \left[  1+(\frac{Z\alpha_{0}}{n-(j+\frac{1}{2})+\sqrt
{(j+\frac{1}{2})^{2}-(Z\alpha_{0})^{2}}})^{2}\right]  ^{-\frac{1}{2}}%
\end{eqnarray}

where $Z\alpha_{0}$ is the product of the charge on the nucleus and the
fine structure constant. \ In the hydrogenic approximation the bound $e^{-}e^{+}$ pair energy
difference would be%
\begin{eqnarray}
\Delta\epsilon(n,j;n',j')&=&\Delta\epsilon(S;S') \nonumber \\
&=&E_{+}(n,j)-E_{-}(n^{\prime},j^{\prime}) \nonumber \\
&=&E_{+}(n,j)+E_{+}(n^{\prime},j^{\prime})
\end{eqnarray}
where in this study the quantum numbers of the negative energy bound states are primed and the atomic states will be labeled by
atomic shell structure notation $S=K,L1,L2,...$.

To get a better idea of the consequences of these new bound electron positron
pairs in the  $D2$ alternative, consider a $Pb$ atom as an illustrative
example. \ This means that $Z=82$ and we can calculate the lowest three
levels:%
\begin{equation}
E_{\pm}(1S_{1/2})=E_{\pm}(K)=\pm409.4\;KeV,
\end{equation}%
\begin{equation}
E_{\pm}(2S_{1/2})=E_{\pm}(L_{1})=\pm484.9\;KeV,
\end{equation}%
\begin{equation}
E_{\pm}(2P_{3/2})=E_{\pm}(L_{2})=\pm487.6\;KeV.
\end{equation}
In the comparisons with experiments below, we will use the $K,L_{1},L_{2}$
notation for these positive energy states and $K^{\prime},L_{1}^{\prime}%
,L_{2}^{\prime}$ for negative energy states. \ The labels for the metastable
bound $e^{-}e^{+}$ pairs can be labeled by the atom, the positive energy
and the negative energy, for example, $Pb:K\rightarrow K^{\prime}.$ \ This
designation would represent the decay of the metastable state consisting of a
positive energy electron in state $K$ and a negative energy hole from state
$K^{\prime}.$ \ The lowest five different possible energies of different metastable states
using the lowest three atomic states for $Pb$ are shown in the table below. \ %
\begin{center}

\begin{tabular}
[c]{|c|c|}\hline
$Transition\;Pb$ & $Energy\;\Delta\epsilon(S;S^{\prime})\;$\\\hline
$Pb:K\rightarrow K^{\prime}$ & $818.8\;KeV$\\\hline
$Pb:K\rightarrow L1^{\prime}$ & $894.3\;KeV$\\\hline
$Pb:K\rightarrow L2^{\prime}$ & $897.0\;KeV$\\\hline
$Pb:L1\rightarrow L1^{\prime}$ & $969.8\;KeV$\\\hline
$Pb:L2\rightarrow L2^{\prime}$ & $975.2\;KeV$\\\hline
\end{tabular}

\end{center}

If these metastable states can be created in heavy ion collisions, then there
should be a variety of discrete energies available for both beam and target
atoms as the metastable states decay. \

In the $D2$ alternative when the two nuclei are close together during a
collision, the positive and the negative energy states will be pulled toward
zero. \ Since there is so much energy around during a
collision, it is expected that many low lying positive energy states will be
empty and electrons can be excited out of the vacuum into these positive
energy states leaving behind a hole in the negative energy states. \ When the
nuclei are near to their closest approach the difference between the positive
and negative energy levels will be its smallest and the probability of
excitation correspondingly greater. \ As the nuclei separate in the scattering
event one or both of the ions could be carrying the bound metastable $e^{-}e^{+}$
 states that are present in the $D2$ alternative. \

\subsection{Decay Modes for bound $e^{-}e^{+}$ metastable states}

Experimentally, one could probe the various decay modes of these metastable
excited states and identify the various states by measuring the energy of the
decay modes. \ In order to do this we first need to examine some of the decay
modes. \

These metastable bound $e^{-}e^{+}$ states can decay in a number of ways.  In order to describe these different channels, we adopt a modified Feynman diagram description in which we denote the ion lines as a broader line if the ion contains a bound $e^{-}e^{+}$ pair and a slightly narrower line if there is no such excitation in the ion.  The electron and positron lines will be thin lines.

The strongest channel for decay of these metastable states would be the annihilation of the bound $e^{-}e^{+}$ pair state and the emission of a single photon.  Because the ion can take up the recoil momentum, a single photon decay is possible.  This is displayed in part $A$ of Fig. 3.  In these diagrams the time moves from left to right and the change in the width of the ion line represents the initial presence and then absence of the metastable state.  This mode of metastable state decay can be either spontaneous, stimulated, or induced by other interactions.   As will be discussed later, at high beam currents, in the presence of many ions, $e^{-}e^{+}$ pairs, and quickly changing fields, it might be expected that the metastable state lifetime is  shortened.

\begin{figure}
\includegraphics[width=\columnwidth]{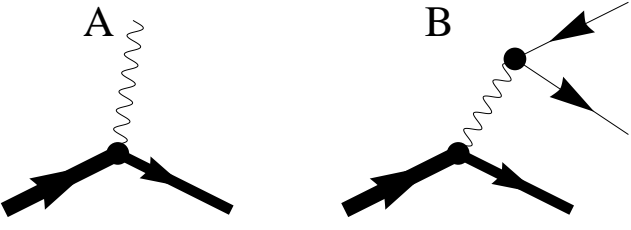}
\caption{Relaxation mechanisms for ion with bound $e^{-}e^{+}$ pair (thick line) and without (thinner line).  A: emission of photon upon annihilation of bound $e^{-}e^{+}$ pair.  B: emission of photon and creation of free $e^{-}e^{+}$ upon annihilation of bound $e^{-}e^{+}$ pair. }
\label{fig:3}
\end{figure}

An example of other decay mechanisms can include the creation of free $e^{-}e^{+}$ pairs.  A lowest order process is shown in part $B$ of Fig. 3.  One would expect this process to give a broad range of pair energies starting at the $2m$ threshold.  This process would not be expected to be a source of the sharply defined free $e^{-}e^{+}$ pair energy distribution.  This process will require quite a large energy change in the ion kinetic energy to reach the pair production threshold.

Another way in which the scattered, metastable ion could create a free $e^{-}e^{+}$ pair would be to join with an available photon.  The diagram for this process is found in part $C$ of Fig. 4.  Here again it is hard to see how this process could give rise to a narrow kinetic energy pair peak.  This process is a twisted form of the Bremsstrahlung scattering in a coulomb field, but with the ion recoil and bound $e^{-}e^{+}$ decay.  Figure 4D shows a process in which the pair energy must come from the change in the ion kinetic energy and the bound pair energy.  It is not expected produce narrow free pair lines.

\begin{figure}
\includegraphics[width=\columnwidth]{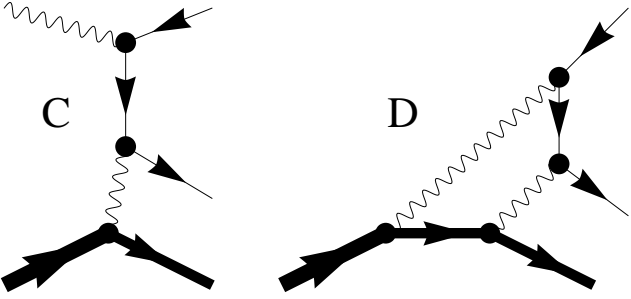}
\caption{Relaxation mechanisms for ion with bound $e^{-}e^{+}$ pair (thick line) and without (thinner line).  C: emission of a free $e^{-}e^{+}$ pair in conjunction with absorption of a photon and annihilation of bound $e^{-}e^{+}$ pair.  D: emission of a free $e^{-}e^{+}$ pair in conjunction with two virtual photons and annihilation of a bound $e^{-}e^{+}$ pair. }
\label{fig:4}
\end{figure}

There is at least one process that combines a possibly sharply delineated total energy with both the threshold $2m$ and the metastable bound state energies $\Delta\epsilon(S,S')$.  It is of sixth order and one of the 16 diagrams is shown in Fig. 5.  There are two vertices at which the excitation energy $\Delta\epsilon(S,S')$ can be transferred and for each of these there are 8 different ways the virtual photons can be arranged.

\begin{figure}
\includegraphics[width=\columnwidth]{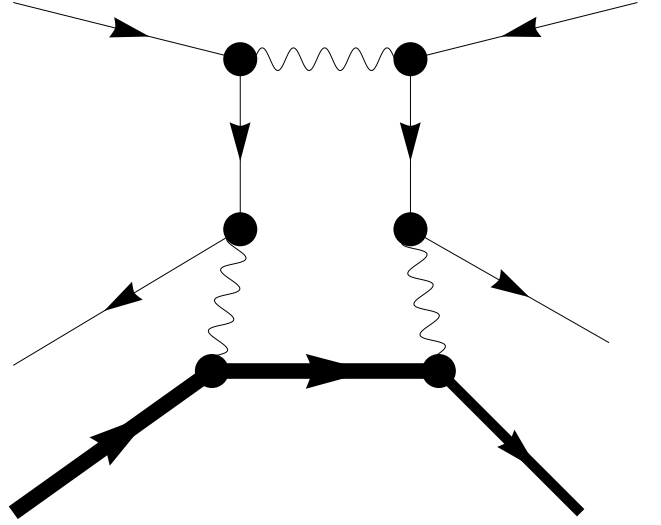}
\caption{Relaxation mechanisms for ion with bound $e^{-}e^{+}$ pair (thick line) and without (thinner line).  Annihilation of a free $e^{-}e^{+}$ pair emission of a free $e^{-}e^{+}$ pair in conjunction with three virtual photons and annihilation of bound $e^{-}e^{+}$ pair.   }
\label{fig:5}
\end{figure}

\subsection{Evaluation of the cross section as a function of energy}

The evaluation of the amplitude and the cross section for this diagram raises
a number of specific issues. The center of mass of the scattered ion can be
treated as a Dirac particle. \ The electron/positron states in the diagram in
lowest order can be treated as free particles. \ The internal degrees of
freedom of the atom must be treated in terms of hydrogenic solutions of the
Dirac equation and in calculating the cross section energy projection
operators for the hydrogenic Dirac equation must be included in the product of
operators. \ The full evaluation of the diagrams will not be
included in this paper. \ Here we are only going to discuss the general structure of this diagram
and examine the energy at which it might be possible to observe sharp lines in
electron positron energies. \

The first such consideration is that we expect the annihilation of the
initial free electron positron pair to be most important for very low
energies. \ It is important to remember that in the limit of small relative
velocities $v_{rel}$ of the electron and positron pair\ the cross section for
pair annihilation blows up at very small $v_{rel}$ and has the form\cite{BandD}%
\begin{equation}
\sigma_{ann}\approx\frac{Const}{v_{rel}}.
\end{equation}

We should expect that the cross section for this process should be able to be
written as%
\begin{equation}
\sigma_{e^{+}e^{-}}=\sum_{{}}\frac{|M^{\prime}|^{2}}{v_{rel}}\delta
(E_{f}-E_{i}),
\end{equation}
where $|M^{\prime}|^{2}$ is the square of the invariant amplitude and all of
the other factors involved in evaluation of this amplitude, $E_{f}$ is the
final state energy, and $E_{i}$ is the initial state of the process. \ The
summation indicates the sums over interior coordinates. \

The final state energy includes the kinetic energy of the final free electron
pair $T_{+}+T_{-}$ , the final kinetic energy of the ion $KE_{f},$ and the
final internal energy of the atom $E(S_{f})$ ($S_{f}$ is the designation of
the atomic electronic states), and $m$ is the electron rest mass.%
\begin{equation}
E_{f}=T_{+}+T_{-}+2mc^{2}+KE_{f}+E(S_{f}).
\end{equation}
The initial energy will depend on the initial kinetic energy of the ion
$KE_{i}$ , the initial internal energy of the atom $E(S_{i}),$ and the kinetic
energy of the initial free electron-positron pair. \ This kinetic energy can
be written for the pair as a center of mass contribution $K_{cm}$ and the
relative kinetic energy $K_{rel}$
\begin{equation}
E_{i}=K_{cm}+K_{rel}+2mc^{2}+KE_{i}+E(S_{i})
\end{equation}
Since we are looking at low energies we consider an average over the relative velocities of the free $e^{-}e^{+}$ states.
In the low energy range of the initial annihilating electron positron pair,
the relative kinetic energy is classical, $K_{rel}=\frac{1}{2} M (v_{rel})^2$, so that an average over the initial
velocities can be written as a sum over the initial relative kinetic energies $K_{rel}$.%
\begin{equation}
\overline{\sigma}=\int D(K_{rel})\frac{|M^{\prime}|^{2}}{\sqrt{K_{rel}}}%
\delta(E_{f}-E_{i})dK_{rel},
\end{equation}
where $D(K_{rel})$ is proportional to the density of states of the low energy
electron positron pairs present during and shortly after the closest approach of the ionic
collision. \ Since $K_{rel}$ must be positive and
\begin{equation}
E_{f}-E_{i}=T_{+}+T_{-}-\Delta\epsilon(S_{i};S_{f})+\Delta KE-K_{cm}-K_{rel}%
\end{equation}
where%
\begin{equation}
\Delta\epsilon(S_{i};S_{f})=E(S_{i})-E(S_{f}),
\end{equation}
and%
\begin{equation}
\Delta KE=KE_{f}-KE_{i},%
\end{equation}
and%
\begin{equation}
\delta\epsilon=\Delta KE-K_{cm},%
\end{equation}

we may write%
\begin{equation}
\overline{\sigma}=\frac{D(x)|M^{\prime}|^{2}\Theta(x)}{\sqrt{x}}.
\end{equation}
where
\begin{equation}
x=T_{+}+T_{-}-\Delta\epsilon(S_{i};S_{f})+\delta\epsilon
\end{equation}
and $\Theta(x)$ is the unit step function. \ This cross section as a
function of the final free electron positron kinetic energy $T_{+}+T_{-}$ has
an infinitely high \ and sharp peak at%
\begin{equation}
T_{+}+T_{-}=\Delta\epsilon(S_{i};S_{f})-\Delta KE + K_{cm}.
\end{equation}

To estimate the lowest possible energy to compare with experiments we will assume
that $K_{cm}$ is negligable.  This assumption restricts our results to values close to
the threshold.  The pairs with non-negligable values will be distributed over a range and will
thus not contribute to a sharp peak.

\subsection{Free $e^{-}e^{+}$ Pairs in the Ion Center of Mass Frame}

The expression for the kinetic energy in the center of mass frame of the system
will be
\begin{equation}
T_{tot}^{cm}=\Delta\epsilon(S;S^{\prime})-\Delta KE_{ion}+K_{cm}.
\end{equation}
\ In what follows we do neglect the kinetic energy $K_{cm}$ of this initial pair because
we seek the lowest energy situation. \

In order to compute the kinetic energy of the free $e^{-}e^{+}$ pair in the lab,
first we must calculate the kinetic energy of the ion in the center of mass
frame. \ Let $v_{I}$ be the magnitude of the lab frame velocity of the ion while it is
in an excited state. \ In the laboratory the ion kinetic energy would be
$\frac{1}{2} M v_{I}^{2}$ ,then the kinetic energy in the center of mass will be
$\frac{1}{2}Mv_{I}^{2}/\gamma_{I}$ where $\gamma_{I}=1/\sqrt{1-\beta_{I}^{2}}%
$, $M$ is the mass of the ion, $\beta_{I}=v_{I}/c,$ and $c$ \ is the velocity
of light.

Initially in the center of mass frame the energy of the ion is
\begin{equation}
E_{1}^{\prime}=2mc^{2}+Mc^{2}+\frac{1}{2}Mv_{I}^{2}/\gamma_{I}+\Delta
\epsilon(S;S^{\prime})
\end{equation}
and the momentum is zero. \

To find the lowest possible energy of the emitted  free $e^{-}e^{+}$ pair, we
consider the momentum change $M\Delta v_{I}$ of the ion to be in the same
direction as the initial velocity and the momentum of the  free $e^{-}e^{+}$
pair will be in the opposite direction and the pair will have an opening angle
of $2\theta_{e}$ . \ The momentum equation in the CM frame for the ion before and after
the emission of the free $e^{-}e^{+}$ pair is
\begin{equation}
0=M\Delta v_{I}-2\gamma_{e}mc\beta_{e}\cos(\theta_{e}).
\end{equation}

The final energy after the emission of the free $e^{-}e^{+}$ pair is%
\begin{equation}
E_{2}^{\prime}=2mc^{2}+\frac{1}{2}M(v_{I}+\Delta v_{I})^2+(\gamma_{e}-1)2mc^{2}.
\end{equation}
Neglecting the term with $(\Delta v_{I})^{2}$ which is very small, we have the
equation%
\begin{equation}
\Delta\epsilon(S;S^{\prime})=(\gamma_{e}-1)2mc^{2}+\frac{M\Delta v_{I}}%
{\gamma_{I}}c\beta_{I}.
\end{equation}
Using the conservation of momentum equation to solve for $M\Delta v_{I}$ and
dividing by $2mc^{2}$ results in
\begin{equation}
\frac{\Delta\epsilon(S;S^{\prime})}{2mc^{2}}=\gamma_{e}-1+\gamma_{e}\beta
_{e}R,
\end{equation}
where
\begin{equation}
R=\frac{\beta_{I}}{\gamma_{I}}\cos(\theta_{e}).
\end{equation}
This results in an equation for $\gamma_{e}$ of the form%
\begin{equation}
\frac{\Delta\epsilon(S;S^{\prime})}{2mc^{2}}+1-\gamma_{e}=R\sqrt{\gamma
_{e}^{2}-1}.
\end{equation}
This equation has two solutions for $\gamma_{e}$%
\begin{eqnarray}
\gamma_{e}^{\pm}&&=\frac{(1+\frac{\Delta\epsilon(S;S^{\prime})}{2mc^{2}}%
)}{1-R^{2}} \nonumber \\
&&\pm\frac{R}{1-R^{2}}\sqrt{\frac{\Delta\epsilon(S;S^{\prime}%
)}{2mc^{2}}(2+\frac{\Delta\epsilon(S;S^{\prime})}{2mc^{2}})+R^{2}}.
\end{eqnarray}

\subsection{Transforming to the laboratory Frame}

In the center of mass frame for the ion, the $e^{-}e^{+}$ pair has the energy
$E_{e^{-}e^{+}}^{\prime}$ and momentum ${P^{\prime}}_{e^{-}e^{+}}$ given by%
\begin{equation}
E_{e^{-}e^{+}}^{\prime}=(\gamma_{e}-1)2mc^{2}+2mc^{2}%
\end{equation}
and%
\begin{equation}
{P^{\prime}}_{e^{-}e^{+}}=-2mc^{2}\gamma_{e}\beta_{e}\cos(\theta_{e}).
\end{equation}
The pair energy $T_{e^{-}e^{+}}^{(lab)}$ in the lab will be given by the Lorentz transformation
\begin{equation}
2mc^{2}+T_{e^{-}e^{+}}^{(lab)}=\gamma_{I}(E_{e^{-}e^{+}}^{\prime}+\beta
_{I}P\prime_{e^{-}e^{+}}).
\end{equation}
From the center of mass energy equation we have%
\begin{equation}
(\gamma_{e}-1)2mc^{2}=\Delta\epsilon(S;S^{\prime})-2mc^{2}\gamma_{e}\beta
_{e}\frac{\beta_{I}}{\gamma_{I}}\cos(\theta_{e})
\end{equation}
and we obtain for the total pair kinetic energy in the laboratory frame%
\begin{eqnarray}
T_{e^{-}e^{+}}^{(lab)}&&=(\gamma_{I}-1)2mc^{2}+\gamma_{I}(\Delta\epsilon
(S;S^{\prime}) \nonumber \\
&&-2mc^{2}\frac{1+\gamma_{I}}{\gamma_{I}}\sqrt{\gamma_{e}^{2}%
-1}\beta_{I}\cos(\theta_{e})).
\end{eqnarray}
This is the expression that should be compared with the experimental peak
locations. \ It depends on the excited state energy of the bound
$e^{-}e^{+}$ pairs, the initial velocity of the scattered ion, and the
opening angle $2\theta_{e}$ of the free $e^{-}e^{+}$ pair in the center of
mass frame of the ion.

The excited bound pairs on the ion are created during the close approach of the
scattering event and the emmision of the free $e^{-}e^{+}$ pairs must occur
after the ion has been scattered. Because most of the spectrometers are
constructed to measure the $e^{-}e^{+}$ pairs perpendicular to the initial
beam direction it suggests that only the ions that have scattered through
small angles will emit $e^{-}e^{+}$ pairs that could be detected. \ Since
there is no knowledge about the scattering angle of the ions in most of the
experiments, we have here assumed that the velocity of the ions after scattering is approximately equal to the magnitude of the initial beam
velocity for all of the experimental situations. \ If the scattering angle of
the ion was known it would be possible to solve for the scattered velocity in
terms of the initial beam velocity and correct this assumption. \ Most of the experiments were conducted
in a range close to $x=6$ Mev per nucleon. \ Accordingly, in the comparison
with experiment, we will use%
\begin{equation}
\gamma_{I}=1+0.001x
\end{equation}
and also determine $\beta_{I}$ from this value for $x=6.$ \

The two solutions for the $\gamma_{e}$ depend on the opening angle
$2\theta_{e}$ of the free pair in the center of mass frame. \ This angle
ranges from $0$ to $\pi/2.$ \ The value of $\theta_{e}=0$ is unphysical since
the two particles would be moving in the same direction on top of each other. \ The value $\theta_{e}%
=\pi/2$ would have the electron and positron moving in opposite directions and
that would allow no recoil of the ion in the center of mass frame. \ The
physical values must be in between. \

\subsection{Comparing Experimental Peak Energies and Free $e^{-}e^{+}$ Lab Energies}

The experimental data has been organized in two tables. \ Table 1 contains
data from spectrometers that measured both the electron and positron energies.
\ Table 2 contains data from earlier experiments in which only the positron
was observed. \ We have first calculated all possible total pair kinetic
energies for each ion (beam and target ) of each experiment. \ \ We initially
chose the value of the transition at $\theta_{e}=\pi/4$ which was closest to
the experimental peak energy. \ This identified the possible transitions that
could have contributed to the peak. \ We then solved for the exact value of
$\theta_{e}$ that would have reproduced that peak. \ In some cases there were
more than one transition possible and we have listed them all. \ If a
transition had a very different value of $\theta_{e}$ it is likely not to be
the correct transition. \ Since there were two different solutions of the
equation for $\gamma_{e}$ each transition will be labeled with the sign of the
solution and the atomic shells. \ For example, one of the smallest transitions
would be labeled $(+)\> U:K\rightarrow K^{\prime}.$ \ Most of the experiments
were well described by shells of $K,L1,L2$ for positive energy states and
$K^{\prime},L1^{\prime},L2^{\prime}$ for negative energy states. \ For some of
the nigher energy peaks it was necessary to use ($M:n=3,j=1/2)$ and for the
very highest transitions $(Z:n\approx50,j=1/2).$ \ Most of the experiments are well
described by the lowest transitions: \ $K\rightarrow K^{\prime},K\rightarrow
L1^{\prime},K\rightarrow L2^{\prime},L1\rightarrow L1^{\prime},L2\rightarrow
L2^{\prime}$ and the peaks were arranged in the corresponding order. \ The
fact, that in the $U+Pb$ experiment we identified some small peaks from the experimental data that would
have never been reported by a good experimentalist, and these fit nicely with
appropriate transitions, is gratifying. \

Each line of the tables includes, besides the experiment identifier, the
following data: \ (1) the experimental peak value, (2) the theory peak value
for  $\theta_{e}=\pi/4,$ (3) the sign ($\pm)$ for the $\gamma_{e}$ solution,
(4) the transition $A:X\rightarrow Y,$ and (5) the value of $\theta_{e}$ which
gives the experimental peak value. \

Every experimental line was identified with at least one plausible transition.  Each of the tables is discussed
in some detail in some remaining paragraphs.

The remaining issue in the heavy ion scattering story historically is the experimental observation
of the APEX collaboration that as the beam current was increased, the sharp $e^{-}e^{+}$ peak slowly
"melted" into the background.  The conclusion that the peaks were thus spurious and unexplainable is based on the assumption
that the peaks represented a stable entity.  However, as seen above, the bound $e^{-}e^{+}$ pair states are
metastable and the stimulated decay rate would certainly increase as the number of scattered ions and other
scattering products increase with beam current.  Appendix C contains arguments that imply the induced decay of the metastable
bound $e^{-}e^{+}$ states will increase with beam current. \ As the induced decay
rate increases, there will be fewer free $e^{-}e^{+}$ pairs to be
observed. \ This metastable state effect could explain why the APEX experiment
failed to see a clear $e^{-}e^{+}$ energy peak as the beam current was increased.

\begin{table}
\caption{ A comparison of the experimental sum-kinetic energy of free $e^{-}e^{+}$ pairs emitted from heavy ions and the theoretical transitions for
specified ions.}
\label{tab:1}
\begin{tabular}
[c]{|c|c|c|c|c|c|}\hline
System,Ref. & Obs.& {Theory($\theta_{e}$=$45^{o}$)} & $\pm$ & Transitions &${\theta_{e}}$\\\hline
U+Pb\cite{Koenig91} & 576 & 521.4 & $+$& U:K$\rightarrow$K'  & $56.0^{o}$\\\hline
 &  & 571 &$+$& Pb:K$\rightarrow$K'  & $46.4^{o}$\\\hline
 & 680* & 677.6 &$-$& Pb:K$\rightarrow$L1'  & $46.0^{o}$\\\hline
 &  & 679.8 &$-$& Pb:K$\rightarrow$L2'  & $45.4^{o}$\\\hline
 &  & 680.9 &$+$& U:L1$\rightarrow$L1'  & $45.3^{o}$\\\hline
 &  & 687.8 &$+$& U:L2$\rightarrow$L2'  & $43.9^{o}$\\\hline
 & 734* & 740.8 &$-$& Pb:L1$\rightarrow$L1'  & $43.5^{o}$\\\hline
 &  & 745.3 &$-$& Pb:L2$\rightarrow$L2'  & $42.4^{o}$\\\hline
 & 787 & 784.5 &$-$& Pb:Z$\rightarrow$Z'  & $69.2^{o}$\\\hline
 & 934* & 784.5 &$-$& U:Z$\rightarrow$Z'  & $69.3^{o}$\\\hline
U+U\cite{Koenig91} & 553 & 563.5 &$-$& U:K$\rightarrow$K'  & $42.0^{o}$\\\hline
 & 634 & 645.2 &$-$& U:K$\rightarrow$L1'  & $42.2^{o}$\\\hline
 & 634 & 649 &$-$& U:K$\rightarrow$L2'  & $41.0^{o}$\\\hline
Th+Th\cite{Th+Th} & 595 & 574.5 &$-$& Th:K$\rightarrow$K'  & $51.1^{o}$\\\hline
 &   & 574.5 &$+$& Th:K$\rightarrow$L1'  & $47.9^{o}$\\\hline
 &608&607.5  &$+$& Th:K$\rightarrow$L1'  & $43.8^{o}$\\\hline
 &   & 610.8 &$+$& Th:K$\rightarrow$L2'  & $44.8^{o}$\\\hline
U+Th\cite{Bokemeyer91}
 & 760$\pm2$ & 763.3 &$-$& Th:M$\rightarrow$M'  & $44.6^{o}$\\\hline
 &  & 762.2 &$-$& U:M$\rightarrow$M'  & $44.8^{o}$\\\hline
U+Ta\cite{Koenig91}\cite{Bokemeyer91} & 630$\pm8$ & 645.1 &$-$& U:K$\rightarrow$L1'
 & $40.9^{o}$\\\hline
 &  & 649 &$-$& U:K$\rightarrow$L2'& $39.8^{o}$\\\hline
 & 746 & 750.9 &$-$& Ta:L1$\rightarrow$L1'  & $44.2^{o}$\\\hline
 & 805$\pm8$ & 784.5 &$-$& Ta:Z$\rightarrow$Z'  & $50.0^{o}$\\\hline
\end{tabular}
\end{table}

\subsection{\ Discussion of Table I}

Table I contains the measured peak locations for a number of the experiments
which have been published in the literature. \ A peak energy with a *
represents a less than significant peak identified by the authors of this theoretical study and would
probably not be regarded as significant by the original authors of the
experiment. Each line of the table includes, besides the experiment identifier, the
following data: \ (1) the experimental peak value, (2) the theory peak value
for  $\theta_{e}=\pi/4,$ (3) the sign ($\pm)$ for the $\gamma_{e}$ solution,
(4) the transition $A:X\rightarrow Y,$ and (5) the value of $\theta_{e}$ which
gives the experimental peak value. It is satisfying that the less significant peaks in the Pb+Pb experiments are fit so well
to the low lying level of the Pb atoms.

\begin{table}
\caption{ A comparison of experimentally measured free positron energies and
predicted energies from designated transitions for
specific ions. \ The theoretical positron energy was assumed to be half of the
total kinetic energy of the pair.}
\label{tab:2}
\begin{tabular}
[c]{|c|c|c|c|c|c|}\hline
System,Ref.&Obs.$E_{+}$&$E_{+}$($\theta_{e}=45^{o}$)&$\pm$& Transition & $\theta_e$\\\hline
U+Cm(epos) & 328$\pm9$ & 322.6,324.5 &-& U:K$\rightarrow$L1',L2'& $48,47$\\\hline
& 328$\pm9$ & 335.5 &+& Cm:L1$\rightarrow$L1'& $42.6$\\\hline
U+Cm(epos) & 445$\pm12$ & 392.2 &-& Cm:Z$\rightarrow$Z'& $67$\\\hline
Th+Cm(epos) & 354$\pm10$ & 365.4 &-& Th:L1$\rightarrow$L1'& $39.3$\\\hline
 & 354$\pm10$ & 358.7 &-& Cm:L1$\rightarrow$L1'& $42.9$\\\hline
Th+Cm(epos) & 367$\pm9$ & 363.9 &-& Cm:L2$\rightarrow$L2'& $46.9$\\\hline
Th+Cm(epos) & 420$\pm11$ & 392.2 &-& Cm,Th:Z$\rightarrow$Z'& $57.6$\\\hline
U+U(orange) & 283 & 281.7 &-& U:K$\rightarrow$K'& $46.1$\\\hline
U+U(epos) & 354 & 363.6 &-& U:K$\rightarrow$K'& $40.2$\\\hline
Th+U(epos) & 367 & 368.5 &-& Th, U:L2$\rightarrow$L2'& $45.2$\\\hline
U+Th(orange) & 291 & 281.8 &-& U:K$\rightarrow$K'& $50.7$\\\hline
 & 291 & 287.3 &-& Th:K$\rightarrow$K'& $47.6$\\\hline
U+Th(epos) & 354 & 344.9 &-& Th:L2$\rightarrow$L2'& $48.5$\\\hline
U+Th(epos) & 459 & 392.2 &-& U, Th:Z$\rightarrow$Z'& $71.9$\\\hline
Th+Th(epos) & 314 & 326.1,327.8 &-& Th:K$\rightarrow$L1',L2'& $38$,$37$\\\hline
Th+Ta(epos) & 367 & 368.5 &-& Th:L2$\rightarrow$L2'& $44.6$\\\hline
 & 367 & 375.5 &-& Ta:L1$\rightarrow$L1'& $40.9$\\\hline
U+Au(orange) & 261 & 260.7 &+& U:K$\rightarrow$K'& $45.5$\\\hline
 & 261 & 281.8 &-& U:K$\rightarrow$K'  & $30.4$\\\hline
U+Au(orange) & 327 & 320.3,321.2 &+& Au:K$\rightarrow$L1',L2'& $48$,$47$\\\hline
& 327 & 322.6,324.5 &-& U:K$\rightarrow$L1',L2'  & $48$,$47$\\\hline
Pb+Pb(orange) & 331 & 338.8,339.9 &-& Pb:K$\rightarrow$L1',L2'& $41$,$40$\\\hline
U+Ta(orange) & 302 & 304.2 &+& Ta:K$\rightarrow$K'& $44.3$\\\hline
 & 302 & 300.3,302.2 &+& Ta:U$\rightarrow$L1',L2'& $46,45$\\\hline

\end{tabular}
\end{table}

\subsection{Discussion of Table II}

The experimental efforts to study these effects were carried out principally
by two groups whose spectrometers were given the labels:\ EPOS and ORANGE.
\ In the early manifestations of these spectrometers only the positron energy
could be determined. \ Much of the early knowledge of these effects was
obtained by observing only the positron energies. \ In later years both
spectrometers gained the capability of
simultaneously observing the positron and electron pairs. \ Both spectrometers
were constructed assuming that the entity emitting the electron positron pairs
had its initial momentum along the beam axis. \ In this case the pairs should
have been observed in coincidence. \ However, it was observed, especially by
the EPOS group that there could be quite significant delays in the arrival of
both the positron and electron. \ It is now clear that a significant cause of
this could be the fact that the scattered ions could be not moving along the original
beam direction when the free electron positron pairs were emitted.

Table II gives the fit of theory to experiment for the positron energies only.  For this comparison the
application of the theory is very simple, the positron energy is assumed to be
half of the energy of the pair. The comparisons are attempted only for the
major peaks observed by experiment. Each line of the table includes: (1) the experiment identifier,
\ (2) the experimental peak value, (3) the theory peak value
for  $\theta_{e}=\pi/4,$ (4) the sign ($\pm)$ for the $\gamma_{e}$ solution,
(5) the transition $A:X\rightarrow Y,$ and (6) the value of $\theta_{e}$ which
gives the experimental peak value.  IN some entries n Table II, where the energies are very close, two cases are listed in one row.  For example, in the first entry $U+Cm$ has two transitions U:K$\rightarrow$L1' andU:K$\rightarrow$L2' and these are combined to read as U:K$\rightarrow$L1',L2' and the corresponding energies and opening half angles are listed with a comma.  For some of the higher energy transitions the energy levels would have to be higher than $K,L1,L2$ and are referred to as Z$\rightarrow$Z'.  In these situations the values are not very different and the transitions are listed as, for example, U, Th:Z$\rightarrow$Z' to indicate that either ion could be involved.

\section{Summary}

The $D2$ alternative, the "path not taken" by Dirac and others, appears to
correct a number of inconsistencies that appeared as a result of the $D1$
alternative. \ In the $D2$ alternative, the reflection symmetry about zero
energy is preserved in the presence of potentials. \ The Dirac equation
including interactions with a vector potential is shown to have charge
conjugation invariance. \ Appropriate Ehrenfest theorems are recovered. \ The
minimal coupling substitution for the $D2$ alternative is found to be%
\begin{equation}
p^{\mu}\rightarrow p^{\mu}-sgn(E) e A^{\mu}(x,t),
\end{equation}
here as in the rest of the paper, SI units are being used.
Some surprising results were found for constant potentials in the $D2$
alternative. \ Negative potentials for positive energy states cannot be deeper
than $-m.$ \ Positive energy and negative energy levels cannot cross the zero
energy axis. \

Positive energy potentials $V_{0}$ that could make up barriers do not
have any upper limit on $V_{0}.$ \ The allowed energy for a wave function is
only for $E>V_{0}+m$ for positive energies and $E<-(V_{0}+m)$ for negative
energy states. \ There are no propagating wave states allowed in the interval $(-V_{0},V_{0}).$
\ The absence of states below the barrier height $V_{0}$ completely suppresses
the Klein tunneling paradox. \ In the $D2$ alternative the physical
interpretation that high static barriers or potential wells induce particle hole
production is no longer necessary.

There is no Klein paradox or tunneling in the $D2$ alternative. As already indicated by
Dragoman\cite{Dragoman}, the usual interpretation that the Klein paradox for a static
potential indicates the production of electron positron pairs is not needed theoretically
in $D1$.  Neither is it needed in $D2$.  The more troubling Klein tunneling is excluded in
alternative $D2$ and so there is little theoretical support for the interpretation of the
Klein paradox as evidence for pair production by static potentials.  Pair production and
annihilation due to dynamical effects obviously remains an important part of relativistic dynamics.

In $D2$ the fact that the charge current density is proportional to the momentum completely replaces the $D1$ requirement
that negative energy states were necessarily part of any wave function, that confinement implies greater negative energy components, and that the current density exhibits zitterbewegung.  All of these $D1$ properties are logically unnecessary in $D2$ and thus
unnecessary as part of the relativistic theory.

There is strong evidence that the observations of free $e^{-}e^{+}$ pairs in
heavy ion scattering experiments can be rationalized in the $D2$ alternative.

The possibility of pulling positive energy states down into the vacuum is forbidden by the symmetry induced
by the charge conjugation invariance of the $D2$ alternative.  This is, at least, consistent with the experimental failure to find unstable vacua in the heavy ion scattering program.

Finally, there are probably many more questions that remain to be explored about the $D2$ alternative.  The discussion
here has been very preliminary, but promising.  Our preliminary stance is that
it appears that Dirac and the physics community should have taken the $D2$
alternative in the first place. \ And that could have made all the difference\cite{pathnottaken}.

\begin{acknowledgements}
The authors would like to acknowledge the useful conversations about this
problem with Mr. John Gray and professor William Kerr. Detailed comments from Prof. J. Reinhardt were extremely helpful.\ The generous
hospitality and detailed discussions of professors Rainer Grobe and Charles
Su during and after a sabbatical (spb) are gratefully acknowledged.
\end{acknowledgements}
\appendix
\section{ Derivation of the Lorentz Force Law}

In this appendix the equation of motion for the generalized momentum is
calculated for the Dirac Hamiltonian in Alternative $D1.$ \ This derivation
follows closely the treatment by Merzbacher\cite{Merzbacher} except in these equations the charge $e$ is the actual charge of the electron, that is, $e<0$. \ The
Hamiltonian is%
$$
H=c\bm{\alpha}\cdot(\bm{p}-\mit e\bm{A})-\mit e\Phi+\beta mc^{2}.\eqno{(A-1)}%
$$
Evaluating the commutators in the equation of motion yields%
$$
\frac{d}{dt}(\bm{p}-\mit e\bm{A})=+e(\bm{E}%
+\bm{v}\times\bm{B}),\eqno{(A-2)}%
$$
as expected. \ In order to re-write this for a more direct comparison with
classical expressions the following identity is used%
$$
\bm{\alpha}\mit (H-e\Phi)+\mit (H-e\Phi)\bm{\alpha}=2c(\bm{p}%
-\mit e\bm{A}),\eqno{(A-3)}%
$$
then the equation becomes%
$$
\frac{d}{dt}(\frac{1}{2c^{2}}(\bm{v}\mit (H+e\Phi)+(H+e\Phi)\bm{v}%
))=-e(\bm{E}+\bm{v}\times\bm{B}).\eqno{(A-4)}%
$$

If the expectation values of this expression are taken to recover the non-relativistic
Ehrenfest classical equations of motion, the expectation values must be taken
with respect to some states of energy either positive or negative. \ In this
case we need to substitute for the following%
$$
\langle H-e\Phi\rangle\approx sgn(E)mc^{2},\eqno{(A-5)}%
$$
where $E$ is the energy of the state, into the equation of motion, yielding%
$$
sgn(E)\frac{d}{dt}(\langle m\bm{v\rangle})=e(\bm{E}%
+\bm{v}\times\bm{B}).\eqno{(A-6)}%
$$
This conclusion within $D1$ yields a classical expression for the momentum,
but the Lorentz force law direction depends on the sign of the energy of the
state determining the expectation value. \

Within the alternative $D2$ the discussion begins with the $D2$ minimal
coupling substitution%
$$
p^{\mu}\rightarrow p^{\mu}-\mit sgn(E)eA^{\mu}\eqno{(A-7)}%
$$
which will generate a new factor of $sgn(E)$ on the right hand side which will
cancel out the same factor on the left hand side, recovering the appropriate
Ehrenfest theorem for the classical limit%
$$
\frac{d}{dt}(\langle \mit m\bm{v\rangle})=\mit e(\bm{E}+\bm{v}%
\times\bm{B}).\eqno{(A-8)}%
$$
This is exactly the Lorentz equation for particle with charge $e$.

\section{ Derivation of the Larmor Spin Dynamics}

The role of the spin in the alternative $D1$ is demonstrated by the equation
of motion of the Matrix operator $\Sigma$ where%
$$
\Sigma=%
\biggl\lbrack
\begin{array}{cc}
\sigma & 0\nonumber\\
0 & \sigma\nonumber
\end{array}
\biggr\rbrack
,\eqno{(B-1)}%
$$
and $\sigma$ are the Pauli Spin matrices. \ This discussion again follows
closely from Merzbacher\cite{Merzbacher}. \ The equation of motion is now for
a Hamiltonian with only a vector potential%
$$
H=c\bm{\alpha}\cdot(\bm{p}-\mit e\bm{A})+\beta
\mit mc^{2}\eqno{(B-2)}%
$$
and proceeds from%
$$
\frac{d\bm{\Sigma}}{dt}=\frac{1}{i\hbar}[\bm{\Sigma}%
,\mit H]=\mit \frac{2c}{\hbar}(\bm{p}-\mit e\bm{A})\times
\bm{\alpha.}\eqno{(B-3)}%
$$
In preparation for taking expectation values with Hamiltonian eigenstates,
consider%
$$
H\frac{d\bm{\Sigma}}{dt}+\frac{d\bm{\Sigma}}{dt}%
H=-2ec\bm{\Sigma\times B.}\eqno{(B-4)}%
$$
In the lowest order expectation value we can substitute $H=sgn(E)mc^{2}$ and
this yields for the expectation values%
$$
sgn(E)\frac{d\langle\bm{\Sigma\rangle}}{dt}=-\frac{e}{mc}%
\langle\bm{\Sigma\rangle\times B,}\eqno{(B-5)}%
$$
which again has the dependence on the sign of energy of the eigenstates used
in the expectation values. \

In the alternative $D2$ the magnetic field will end up with a factor of the
$sgn(E)$ on the right hand side and this will recover the classical Larmor
equation for the spin in a magnetic field%
$$
\frac{d\langle\bm{\Sigma\rangle}}{dt}=-\frac{e}{mc}\langle
\bm{\Sigma\rangle\times B.}\eqno{(B-6)}%
$$

\section{ Metastable State Decay Effect on Experiments}

The decay of these metastable excited ions after the collision with the target
can be studied approximately by evaluating Feynman diagrams. \ There are at
least three processes that are likely to be most important. \ The first of these
is the rate $R_{\phi}$ of spontaneous emission of a photon as the occupied
positive energy electron drops into the empty negative energy state. \ The
second rate $R_{ep}$ is the absorption of an initial free $e^{-}e^{+}$ pair and the decay of the bound $e^{-}e^{+}$ state into a free
electron -positron pair. \ This free $e^{-}e^{+}$ pair is what was detected in
most of the heavy ion scattering experiments. \ A third rate $R_{ind}$ would
be the induced decay of these bound $e^{-}e^{+}$ pairs because of other ions, pairs, photons, and time varying electric fields
in the scattering region after the target. \ This induced decay should be proportional in lowest
approximation to the square of the current since it reflects interactions between beam ions and
their associated post scattering constituents. \ The other two rates should be
proportional to the beam current. \ This means that the production of free
$e^{-}e^{+}$ pairs should be proportional to%

\[
\frac{R_{ep}}{R_{\phi}+R_{ep}+R_{ind}}%
\]

Since every term in this ratio is at least proportional to the beam current,
we can factor out one power of this current and get an expression for the
production cross section for $e^{-}e^{+}$ pairs that has the following structure.%

$$
\sigma_{ep}=\frac{\sigma_{ep}^{0}}{x_{o}+\eta J}\eqno{(C-1)}%
$$

where $\eta J=R_{ind}/$ $R_{ep}$ is a parameter expressing the ratio of the
induced emission term divided by the spontaneous emission term, $J$ is the
beam current, and $x_{o}=(R_{\phi}+R_{ep})/R_{ep}$ and is independent of the
beam current.

Let us now examine the effect of this expression on the time it takes to
collect a statistically significant signal in a scattering experiment. In
carrying out this analysis, let us express the cross section for the
background $e^{-}e^{+}$ pairs as $\sigma_{2}$ . In order to establish a
statistically significant signal we need the ratio of the signal over the
standard deviation to be some predetermined number, $z$ , which can be
expressed as%

$$
z=\frac{\sigma_{ep}Jt}{\sqrt{\sigma_{2}Jt}}\eqno{(C-2)}%
$$

where t is the time the counting experiment must run to get $z$ . In the usual
scattering experiment (essentially the assumption behind the APEX papers),
there is no factor reflecting competing channels for the supply of the
metastable states that give rise to the $e^{-}e^{+}$ pairs, and we can replace
$\sigma_{ep}$ by $\sigma_{ep}^{0}$ and get the following expression for the
time $t$ that it takes to achieve statistical significance.%

$$
t=\frac{\eta z^{2}\sigma_{2}/(\sigma_{ep}^{0})^{2}}{\eta J}=\frac{t_{0}}%
{x}\eqno{(C-3)}%
$$

where for the following comparison we have expressed the formula in
dimensionless variables $x=\eta J$ and $\tau=t/t_{0}$. If we now write in
these units the standard expectation for an experiment, we get%

$$
\tau=\frac{1}{x},\eqno{(C-4)}%
$$
which reflects the experimental wisdom that increasing the beam current
decreases the counting time to achieve statistically significant results.

However, the same expression for the case where metastable bound $e^{-}e^{+}$
states are competing with the spontaneous and induced emission of photons
yields the following form for statistically significant observation times :%

$$
\tau=\frac{(x_{o}+x)^{2}}{x}.\eqno{(C-5)}%
$$
As can be seen by the functional dependence on $x$ this function has a minimum
at $x_{o}$ and then increases as the beam current ($x)$ increases. \ In this
case, large currents will actually increase the counting time, and there is an
optimal range of beam currents which gives the shortest counting time for
observing the emitted free electron positron pairs from the decay of the
metastable states.

Two very different behaviors are to be expected for the counting time to
significance. \ If there is no metastable states that can be induced to decay,
increasing the beam current, always decreases the counting time. \ However, in
the presence of a metastable, inducible, intermediate state, there is an
optimal counting time and increasing the beam current increases the counting
time considerably and can appear to overwhelm the peaks if not enough time is
used for counting. \

%

\end{document}